\documentclass[10pt,journal]{IEEEtran}
\usepackage{amsthm, bm, graphicx, hyperref}
\usepackage{algorithm}
\usepackage{algorithmic}
\usepackage{array}
\usepackage{url}
\usepackage{cite}
\usepackage{autobreak}
\usepackage{amssymb}
\usepackage{subfigure}
\usepackage{caption}
\usepackage{xcolor}

\newtheorem{theorem}{\bf Theorem}
\newtheorem{lemma}{\bf Lemma}

\newtheorem{mypro}{Proposition}
\newcommand{\bs}[1]{\boldsymbol{#1}}

\newcommand{\ib}[1]{\in\mathbb{#1}}
\newcommand{\ic}[1]{\in\mathcal{#1}}

\begin{document}
\title{Movable Antenna-Enhanced Near-Field Flexible Beamforming: Performance Analysis and Optimization}

\author{
Shun~Yang,
Xin~Wei,
Nianbing~Su,
Weidong~Mei,~\IEEEmembership{Member,~IEEE,}
Zhi~Chen,~\IEEEmembership{Senior Member,~IEEE}\\ 
and Boyu~Ning,~\IEEEmembership{Member,~IEEE} 

\thanks{S. Yang, X. Wei, W. Mei, Z. Chen and B. Ning are with the National Key Laboratory of Wireless Communications, University of Electronic Science and Technology of China, Chengdu 611731, China (e-mails: shunyang@std.uestc.edu.cn, xinwei@std.uestc.edu.cn, wmei@uestc.edu.cn, chenzhi@uestc.edu.cn, boydning@outlook.com).}%
\thanks{N. Su is with the Glasgow College, University of Electronic Science and Technology of China, Chengdu 611731, China (e-mails: nianbingsu@std.uestc.edu.cn). He is also with the School of Engineering, University of Glasgow, Glasgow G12 8QQ, U.K.}
} 
\maketitle

\begin{abstract}
 As an emerging wireless communication technology, movable antennas (MAs) offer the ability to adjust the spatial correlation of steering vectors, enabling more flexible beamforming compared to fixed-position antennas (FPAs). In this paper, we investigate the use of MAs for two typical near-field beamforming scenarios: {\bf{beam nulling}} and {\bf{multi-beam forming}}. In the first scenario, we aim to jointly optimize the positions of multiple MAs and the beamforming vector to maximize the beam gain toward a desired direction while nulling interference toward multiple undesired directions. In the second scenario, the objective is to maximize the minimum beam gain among all the above directions. However, both problems are non-convex and challenging to solve optimally. To gain insights, we first analyze several special cases and show that, with proper positioning of the MAs, directing the beam toward a specific direction can lead to nulls or full gains in other directions in the two scenarios, respectively. For the general cases, we propose a discrete sampling method and an alternating optimization algorithm to obtain high-quality suboptimal solutions to the two formulated problems. Furthermore, considering the practical limitations in antenna positioning accuracy, we analyze the impact of position errors on the performance of the optimized beamforming and MA positions, by introducing a Taylor series approximation for the near-field beam gain at each target. Numerical results validate our theoretical findings and demonstrate the effectiveness of our proposed algorithms.
\end{abstract}

\section{Introduction}
With the rise of high-frequency technologies such as millimeter-wave and terahertz, electromagnetic waves tend to propagate as spherical rather than planar waves between base stations and users. The resulting wavefront curvature leads to a nonlinear relationship between phase differences and antenna element spacing, rendering it different from far-field propagation. To tackle this non-linearity, existing works have proposed various solutions in terms of beamforming/codebook design and channel estimation dedicated to near-field and hybrid near- and far-field propagation (see e.g., \cite{Liu2023NF,An2024NF,Cong2024NF,Cui2023NF,Ning2023Beamforming} and the references therein).

\begin{figure}[!t]
\centering 
\includegraphics[height=3.8cm]{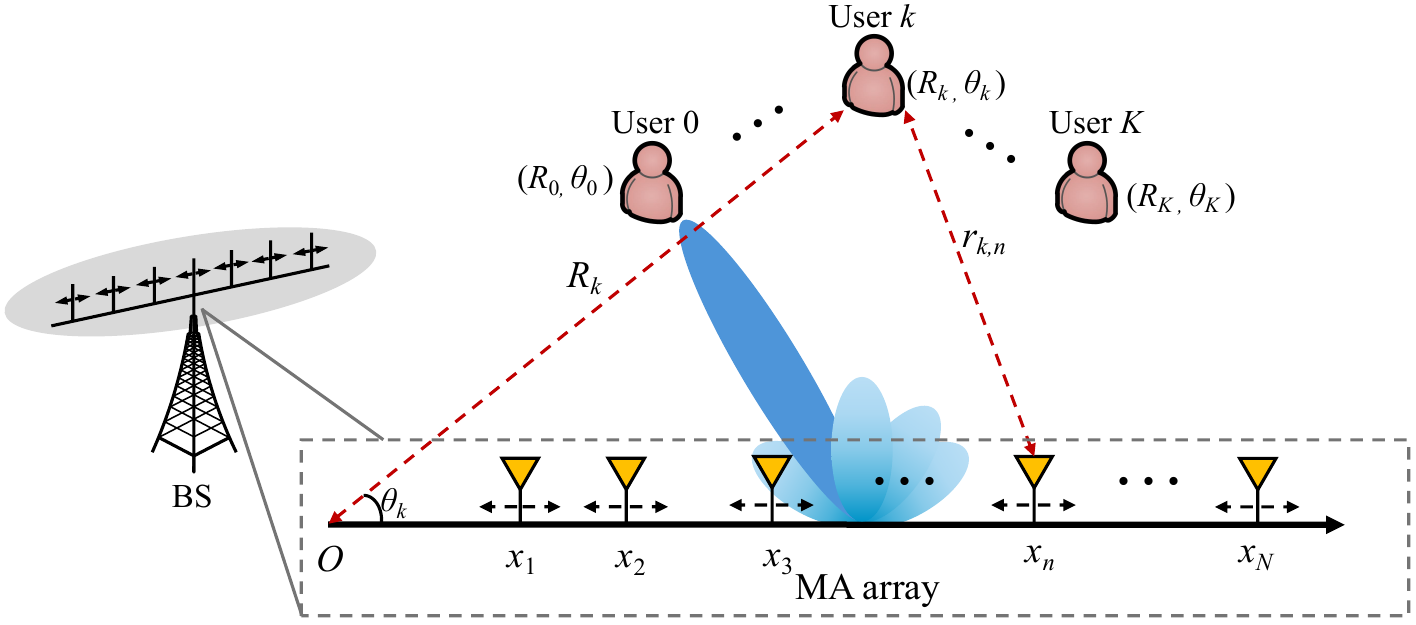}
\caption{MA-enhanced near-field beam nulling (Scenario 1).}
\label{fig_1}
\vspace{-0.1cm}
\end{figure}

\begin{figure}[!t]
\centering
\includegraphics[height=3.8cm]{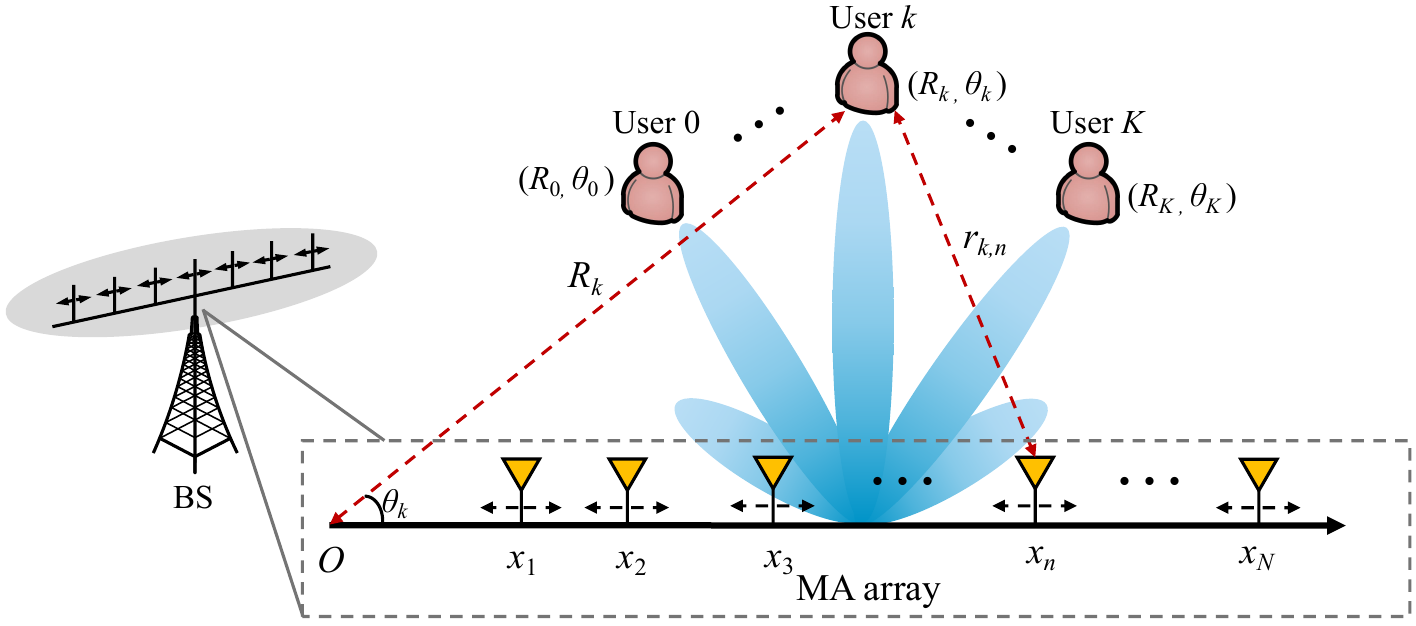}
\caption{MA-enhanced near-field multi-beam forming (Scenario 2).}   
\vspace{-6pt} 
\label{fig_2}
\end{figure}

However, all existing solutions rely on conventional fixed-position antennas (FPAs), where the spatial correlation between steering vectors at any two angles remains constant. This limitation can lead to suboptimal performance, leaving room for further improvement. Fortunately, the movable antenna (MA) technology has aroused significant attention in the area of wireless communications{\cite{LiMA2025,6DMA,li2024sum,ma2025robust,Add1zhu2024mag,Add2zhu2025}}.  Compared to FPAs, MAs introduce a new spatial degree of freedom (DoF) to alter the spatial correlation of steering vectors for different angles by allowing each antenna to flexibly adjust its position within a confined region \cite{Zhou2024MA,Wei2025MARIS,Ning2025MA,Add4Gao2024}. To leverage this advantage, the authors in \cite{zhu2023movable} explored MA-enabled beam nulling by jointly optimizing the antenna weight vector (AWV) and antenna position vector (APV) to maximize beam gain toward a target direction while nulling gains at multiple undesired directions. The results showed that this goal can be achieved by applying maximum-ratio transmission (MRT) toward the target direction with proper antenna positioning. Furthermore, the authors in \cite{ma2024multi} investigated MA-enhanced multi-beam forming to generate high-gain beams toward multiple directions while suppressing interference, through joint AWV and APV optimization. In \cite{ZhuMABF}, MAs were mounted on satellites to address a more general dynamic beam coverage problem, where the antenna geometry was adjusted in real time based on the satellite’s position. Different from the above works focusing on the beam gain at several discrete directions, the authors in \cite{wang2025coverage,wang2025movable} applied MAs for achieving uniform beam coverage in a continuous region in the angular domain. In addition to enabling more flexible beamforming, MAs offer several additional benefits in wireless communications, including signal power enhancement\cite{mei2024movable,Zhu2024MA_WCOM,Yang2024MA,Add3Gao2024}, interference suppression\cite{wei2024joint,xiao2024multiuser,zhu2024movable}, wireless sensing \cite{LyuMA2025Sensing,ma2024movable,shao2024exploiting,jiangMA2025,wang2025antenna} and physical-layer security\cite{hu2024secure,Shen2025MAPhy,mei2024security,Rostami2024Security,Tang2025MASec,Hu2024MASecure}.\\
 \indent Despite the recent results for MA-enhanced flexible beamforming\cite{Feng2024MAFF,Wang2024MABF}, they are mostly derived assuming far-field propagation conditions. In order to address the high demand for beamforming in near-field communication\cite{Wang2024NFBF,NFAp2024,Li2024_NF,Guo2024BFNF}, we focus on MA-enhanced near-field beamforming in this paper in two typical scenarios, namely, beam nulling and multi-beam forming, as illustrated in Figs.\,\ref{fig_1} and \ref{fig_2}, respectively. Notably, due to the significant wavefront curvature, the existing AWV and APV solutions might need to be reconsidered. Besides wavefront curvature, another practical challenge in implementing MA systems lies in the inevitable position errors in antenna position adjustment. These errors stem from factors such as mechanical limitations and calibration inaccuracies. Due to the rapid spatial variation of electromagnetic field properties, even small position errors may lead to significant degradation in beamforming performance, particularly for near-field beamforming relying on both positions and angles. Therefore, an accurate analysis of the impact of position errors is essential for developing robust MA systems, which, however, is still lacking in existing literature even for far-field beamforming with MAs. 


To address the above limitations, this paper conducts performance analysis and optimization for MA-enhanced near-field beamforming, and also captures the effects of MA position errors on the beamforming performance. The main contributions are summarized as follows:
\begin{itemize}
  \item For the beam-nulling scenario, we aim to jointly optimize the antenna position vector (APV) of multiple MAs and the beamforming vector to maximize the beam gain at a desired target (e.g., user 0 in Fig.\,\ref{fig_1}) while nulling interference toward undesired targets (e.g., users 1 to $K$). This joint optimization problem is non-convex and challenging to solve optimally. To gain insights, we first analyze several special cases under the assumption of an arbitrarily large movement region (ALMR) \cite{Zhu2024MA_WCOM}. Our analysis shows that even under near-field propagation, MRT toward user 0 can still achieve interference nulling at other users, which extends the results in \cite{zhu2023movable} to the more complex near-field setup. For the general case, we propose a sequential update algorithm to obtain a high-quality suboptimal APV by discretizing the antenna movement region and applying zero-forcing (ZF) beamforming.
  \item For the multi-beam forming scenario, our goal is to maximize the minimum beam gain among all target directions in Fig.\,\ref{fig_2} by jointly optimizing the APV and transmit beamforming. Under the ALMR assumption, we show that full beam gain, also known as grating lobes, can be approximately achieved at all targets using a uniform antenna array. In the general case, we develop an alternating optimization (AO) algorithm that decouples the problem into two subproblems for beamforming and APV optimization. They are then efficiently solved using successive convex approximation (SCA) and a sequential update method, respectively.
  \item Next, we characterize the impact of antenna position errors on near-field flexible beamforming with MAs. In particular, we aim to derive the worst-case sum beam gain towards these directions under norm-bounded antenna position errors. To deal with the intricate near-field steering vectors, we introduce Taylor expansion to approximate the near-field beam gain toward each user. Based on this, we show that the worst-case (maximum) sum beam gain for undesired directions can be derived by solving a quadratically constrained quadratic programming (QCQP) problem. While in the multi-beam forming scenario, we derive the worst-case (minimum) sum beam gain for all directions in closed-form, revealing key factors that contribute to performance degradation due to antenna position errors. Finally, numerical results are provided to validate the theoretical analysis and demonstrate the effectiveness of the proposed algorithms. The results show that our algorithm outperforms the widely used particle swarm optimization (PSO) method. Moreover, the proposed approximation technique achieves high accuracy in near-field beam gain.
  \end{itemize}
  
\indent The remainder of this paper is organized as follows. Section II presents the system model and problem formulation for the two considered scenarios. Section III presents theoretical analysis of the two scenarios in several special cases. Section IV presents the proposed algorithms for solving the formulated problems in the general case. Section V analyzes the worst-case beam gain in the two scenarios in the presence of antenna position errors. Simulation results are provided in Section VI, and this paper is finally concluded in Section IX.

{\it Notations}: $a$, $\boldsymbol{a}$, $\boldsymbol{A}$, and $\mathcal{A}$ denote a scalar, a vector, a matrix and a set respectively. $(\cdot)^T$, $(\cdot)^H$, and $(\cdot)^{-1}$ denote the transpose, conjugate transpose and inverse of a matrix, respectively. $\mathbb{R}$ and $\mathbb{C}$ denote the sets of real numbers and complex numbers, respectively. $|a|$ and $\|\boldsymbol{a}\|_2$ denote the amplitude of a scalar $a$ and the norm of a vector $\boldsymbol{a}$, respectively. The real and imaginary parts of a complex number are denoted by $\Re\{\cdot\}$ and $\Im\{\cdot\}$, respectively.\vspace{-6pt}

\begingroup
\allowdisplaybreaks
\section{System Model And Problem Formulation}
As shown in Figs.\,\ref{fig_1} and \ref{fig_2}, we consider a BS equipped with a linear array of $N$ MAs, serving $K+1$ users  located in its near field. Given a reference point on the linear MA array, denoted as $O$, let $(R_k, \theta_k)$, $k\in\mathcal{K} \triangleq \{1, 2, \cdots, K\}$ denote the geographic location of user $k$, where $R_k$ represents its distance with the reference point $O$, and $\theta_k\in[0,\pi]$ denotes the angle of departure from the antenna array to user $k$. Let $x_n$ denote the coordinate of the $n$-th MA, $n \in \mathcal{N} \triangleq \{1, 2, \cdots, N\}$. Thus, the APV of the MA array can be expressed as $\boldsymbol{x} = [x_1, x_2, \ldots, x_N]^T \in \mathbb{R}^{N \times 1} $. The distance from the $n$-th MA to  user $k$ is given by
\begin{equation}
r_{k,n}=R_k\sqrt{1+\frac{x_n^2}{R_k^2}-\frac{2x_n}{R_k}\cos\theta_k}. 
\end{equation}
By introducing the Taylor expansion $\sqrt{1+x}\approx1+\frac12x-\frac18x^2$ as in \cite{NFAp2019,NFAp2024}, $r_{k,n}$ can be approximated as
\begin{equation}\label{eqn_Dist}
    r_{k,n} \approx R_{k}-x_{n}\cos\theta_{k}+\frac{x_{n}^{2}}{2R_{k}}\sin^{2}\theta_{k}. 
\end{equation}

\indent In near-field communications, the spherical wavefront model should be considered. As a result, the near-field steering vectors of the MA array can be written as a function of the APV $\boldsymbol{x}$, $\theta_k$ and $R_k$, i.e.,
  \begin{equation}\label{eq3}
     \boldsymbol{a}(\boldsymbol{x},R_{k},\theta_{k})=\left[e^{j\frac{2\pi}{\lambda}(r_{k,1}-R_{k})},\cdots,e^{j\frac{2\pi}{\lambda}(r_{k,N}-R_{k})}\right]^T,
  \end{equation}
where $\lambda$ denotes the carrier wavelength of the considered system.
{The boundary between near-field and far-field propagation conditions is generally characterized by the Rayleigh distance, denoted as $R_{\text{Rayleigh}} = 2D_{\max}^2/\lambda$, where $D_{\max}$ denotes the length of the linear MA array. This indicates that even with a limited number of MAs, the Rayleigh distance can still be large if their movement region is sufficiently large. It is also noted from \eqref{eq3} that unlike far-field steering vectors that depend only on angle, the near-field steering vectors depend on both distance and angle, thus offering an additional degree of freedom for beam steering.} Denoting $\boldsymbol{w}\in\mathbb{C}^{N\times 1}$ as the unit-power AWV for beamforming, i.e., $\lVert \boldsymbol{w} \rVert^{2}_2=1,$ the beam gain at user $k$ can be expressed as
\begin{equation}\label{eqn_BeamGain}
  G(\boldsymbol{x},\boldsymbol{w},\theta_k, R_k) =\left|\boldsymbol{w}^{H}\boldsymbol{a}(\boldsymbol{x},R_{k},\theta_{k})\right|^{2}.
\end{equation}
\indent In this paper, to characterize the performance gain provided by MAs over conventional FPAs, we consider two typical scenarios of MA-enhanced near-field flexible beamforming, i.e., beam nulling and multi-beam forming, as depicted in Figs. 1 and 2, respectively. {In particular, this paper focuses on the beam gain in \eqref{eqn_BeamGain} as the performance metric from an array signal processing perspective.}  {This provides fundamental insights into the beamforming characteristics of MAs versus FPAs, without involving specific user scheduling or resource allocation algorithms. The results can also be practically applied to ensure users’ average performance based on their spatial distribution, i.e., line-of-sight (LoS) channel components under near-field propagation conditions. Since user distribution typically changes slowly over time, the associated statistical CSI, i.e., $(R_k, \theta_k), k \in {\cal K}$, can be obtained offline\footnote{{This also facilitates statistical CSI acquisition for undesired users, regardless of whether they are external or internal.}}, thus avoiding frequent antenna movement that may incur large movement delay and energy consumption. In addition, the transmit beamforming can be re-optimized based on the users’ instantaneous channels to meet their real-time performance requirements.}

\vspace{-6pt}

\subsection{Scenario 1: Beam Nulling}
In the first scenario, the objective is to form a high-gain beam directed at a target user (user 0), while nulling the beam gains towards $K$ undesired users. Thus, we can formulate the following optimization problem in terms of $\bs{x}$ and $\bs{w}$:
\begin{subequations}\label{eqn_OptPrblm_P1}
	\begin{align}
		{\text{(P1)}}\quad & \underset{\boldsymbol{x},\boldsymbol{w}}{\max}\quad G(\boldsymbol{x},\boldsymbol{w},\theta_0, R_0)\\
   		\mathrm{s.t.}\quad  &x_{1}\geq0,x_{N}\leq D_{\max}, \label{eqn_Cons_Region}\\
    	& \left|x_{n}-x_{n-1} \right| \geq D_{\min}, \ n=2,3,\ldots,N,\label{eqn_Cons_MinDist}\\
    	&G(\boldsymbol{x}, \boldsymbol{w},\theta_k, R_k) = 0, \ k=1,2,\ldots,K, \label{eqn_BeamGain_4d}\\
    	& \|\boldsymbol{w}\|_{2}\leq1,\label{eqn_Cons_Power4e}
    \end{align}
    \end{subequations}
    where constraint $\eqref{eqn_Cons_Region}$ ensures that the MAs are moved within the MA array; {constraint $\eqref{eqn_Cons_MinDist}$ ensures a minimum inter-antenna spacing, i.e., $D_{\min}$, to avoid mutual coupling}; constraint $\eqref{eqn_BeamGain_4d}$ ensures the beam-nulling conditions at the $K$ users; and constraint $\eqref{eqn_Cons_Power4e}$ is the power constraint on the transmit beamforming.

Note that for any given APV $\boldsymbol{x}$, the optimal AWV $\boldsymbol{w}$ for (P1) can be obtained as the ZF beamforming, i.e.,
\begin{equation}\label{eqn_ZF}
 \boldsymbol{w}_{\boldsymbol{x}}=\boldsymbol{P}(\boldsymbol{x}) \boldsymbol{a}(\boldsymbol{x}, R_0, \theta_0)
\end{equation}
  where $\boldsymbol{P}(\boldsymbol{x})=\boldsymbol{I}_N - \boldsymbol{A}(\boldsymbol{x}) \left( \boldsymbol{A}(\boldsymbol{x})^H \boldsymbol{A}(\boldsymbol{x}) \right)^{-1} \boldsymbol{A}(x)^H$ denotes a projection matrix of any vector into the zero space of $\boldsymbol{A}(\boldsymbol{x})$,  with $\boldsymbol{A}(\boldsymbol{x}) = \left[\boldsymbol{a}(\boldsymbol{x}, R_1,\theta_1), \boldsymbol{a}(\boldsymbol{x},R_2, \theta_2), \ldots, \boldsymbol{a}(\boldsymbol{x},R_K, \theta_K) \right].$\\
  \indent By substituting $\eqref{eqn_ZF}$ into $\eqref{eqn_BeamGain}$, the beam gain at the target user 0 becomes
\begin{equation}\label{eqn_BeamGain_ZF}
    G(\boldsymbol{x}, \theta_0, R_0)=\left| \boldsymbol{a}^H(\boldsymbol{x}, R_0, \theta_0) \boldsymbol{w}_{\boldsymbol{x}}^{\text{ZF}} \right|^2 = N - I(\bs{x}),
\end{equation}
where
\begin{align}\label{EqI(x)}
   I(\bs{x})&\triangleq\bs{a}(\bs{x}, R_0, \theta_0)^H \bs{A}(\boldsymbol{x}) \left( \bs{A}(\bs{x})^H \bs{A}(\bs{x}) \right)^{-1}\bs{A}(\bs{x})^H \nonumber\\
   &\quad\times\bs{a}(\bs{x}, R_0 ,\theta_0).
  \end{align}

\indent As such, problem (P1) can be simplified as
\begin{equation}\label{eqn_P2}
	{\text{(P2)}}\quad \underset{\boldsymbol{x}}{\max}\quad G(\boldsymbol{x}, \theta_0, R_0),\quad\mathrm{s.t.}\quad\eqref{eqn_Cons_Region},\eqref{eqn_Cons_MinDist},
\end{equation}
which is only related to the APV $\boldsymbol{x}$. 

However, due to the intractable form of \eqref{EqI(x)} and the non-convex constraint \eqref{eqn_Cons_MinDist} w.r.t. $\boldsymbol{x}$, problem (P2) is a non-convex optimization problem that is challenging to be optimally solved. In Section III, we first conduct theoretical analyses to show that (P2) can be optimally solved in several special cases. In Section V, we propose an efficient algorithm to obtain a high-quality suboptimal solution to (P2) in the general case.\vspace{-6pt}

\subsection{Scenario 2: Multi-Beam Forming}
In the second scenario, we aim to generate multiple high-gain beams at all $K+1$ users simultaneously. To this end, our goal is to maximize the minimum beamforming gain over $\left\{ (R_k, \theta_k) \right\}_{k=0}^K$, by jointly optimizing the APV $\boldsymbol{x}$ and the AWV $\boldsymbol{w}$. The associated optimization problem can be formulated as 
\begin{subequations}\label{eqn_OptPrblm_P2}
\begin{align}
{\text{(P3)}}\quad &\underset{\boldsymbol{x},\boldsymbol{w},\delta} {\max}\quad  \delta\label{eqn_BeamGain_GL_Objectfun}\\
\mathrm{s.t.}\quad  & G(\boldsymbol{x}, \boldsymbol{w}, \theta_k, R_k)\geq\delta, \ k=0,1,\ldots,K, \label{eqn_BeamGain_GL}\\
& \|\boldsymbol{w}\|_{2}\leq1, \label{eqn_Cons_Power}\\
&\eqref{eqn_Cons_Region},\eqref{eqn_Cons_MinDist},\nonumber
\end{align}
\end{subequations}
where $\delta$ is an auxiliary optimization variable. Note that problem (P3) is a non-convex optimization problem due to the intricate coupling between $\boldsymbol{w}$ and $\boldsymbol{x}$ in constraint $\eqref{eqn_BeamGain_GL}$ and the non-convex constraint (4c). In Sections IV and VI, we derive the optimal solution and high-quality suboptimal solution to (P3) under several special cases and the general case, respectively.\vspace{-6pt}

\section{Special Case Analysis}\label{sc-bn}
In this section, we show that a full beam gain can be achieved at the target user, i.e., $G(\bs{x},\theta_0,R_0)=N$, while nulls and full beam gains can be achieved at the other $K$ users under certain conditions. To facilitate our analysis, we assume the size of the antenna array, i.e., $D_{\max}$, is sufficiently large such that the constraints in \eqref{eqn_Cons_Region} can be relaxed\cite{zhu2023movable,Zhu2024MA_WCOM,wei2024joint}. \footnote{{It is worth noting that the actual movement region of the MAs is always finite. The purpose of assuming an ALMR and conducting asymptotic analysis here is to reveal the stronger beam-steering capability of MAs compared to FPAs, similarly to that for massive multiple-input multiple-output (MIMO)\cite{Björnson2018MIMO} and/or intelligent reflecting surface (IRS)-aided systems\cite{Wu2019IRS}.}} {However, even under the ALMR assumption, problems (P2) and (P3) remain intractable due to their complex beamforming objectives, which give rise to transcendental equations.}\vspace{-6pt}

\subsection{Scenario 1: Beam Nulling}
Note that by setting the beamforming vector as the MRT toward user 0, i.e., $\bs{w}= \bs{a}(\bs{x},R_0, \theta_0)/\sqrt{N}$, the objective function of (P2) can be maximized. Next, we aim to show the existence of an APV $\bs{x}$, such that
\begin{equation}\label{eqn_SV}
f_k(\bs{x})\triangleq\bs{a}\left(\bs{x},R_0,\theta_0\right)^H\bs{a}\left(\bs{x},R_k,\theta_k\right)=0,\forall k\ic{K},
\end{equation}
i.e., {{the beam gains at the $K$ undesired users can be nulled by the MRT toward user 0 with an ALMR.}} For convenience, we first define
\begin{align}
	&a_{k}\triangleq\cos\theta_{0}-\cos\theta_{k},\forall k\ic{K},\\ &b_{k}\triangleq-\left(\frac{\sin^{2}\theta_{0}}{2R_{0}}-\frac{\sin^{2}\theta_{k}}{2R_{k}}\right), \forall k\ic{K},
\end{align}
and start from a simplified case of $K=1$.

\begin{mypro}
If $K=1$, there must exist an APV $\bs{x}$ that satisfies $f_1(\bs{x})=0$ for any given $N\ge 2$.
\end{mypro}
\begin{IEEEproof}
	For $K=1$, $f_1(\bs{x})$ can be expressed as
	\begin{equation}
		f_1(\bs{x})=\sum_{n=1}^{N}e^{j\frac{2\pi}{\lambda}\left(r_{1,n}-r_{0,n}\right)}=\sum_{n=1}^{N}e^{j\frac{2\pi}{\lambda}\left(a_1x_n+b_1x_n^2\right)}.
	\end{equation}

 To construct an APV $\bs{x}$ satisfying $f_1(\bs{x})=0$, the coordinate of the $n$-th MA should satisfy
	\begin{equation}\label{eqn_Lem1_Eq1}
		\frac{2\pi}{\lambda}\left(a_1x_{n}+b_1x_{n}^{2}\right)=\frac{2\pi n}{N}+2\pi q_{n}, \forall n\in\mathcal{N},
	\end{equation}
	where $q_n$, $n\ic{N}$ are integers that ensure constraints \eqref{eqn_Cons_MinDist}. {The condition (15) ensures that the phase terms $\exp(j\frac{2\pi}{\lambda}(a_1 x_n + b_1 x_n^2)), n \in {\cal N}$, are uniformly distributed on a unit circle, leading to perfect cancellation when they are summed over $n$. This geometric interpretation highlights the role of symmetric phase alignment in achieving beam nulling.} Note that \eqref{eqn_Lem1_Eq1} includes a set of quadratic equations, and we can easily obtain the solutions to them as
	\begin{equation}\label{eqn_OptAPV_Lem1}
		x_n=-\frac{a_1}{2b_1}+\sqrt{\frac{a_1^2}{4b_1^2}+\frac{\lambda}{b_1}\left(\frac{n}{N}+q_n\right)},\forall n\in\mathcal{N}.
	\end{equation}
    This thus completes the proof.
\end{IEEEproof}

Proposition 1 indicates that a full beam gain at $(R_0,\theta_0)$ and nulling at $(R_1,\theta_1)$ can be simultaneously achieved in the near-field scenario by optimizing the APV for any given $N\ge2$. In addition, it is worth noting that as the signal propagation environment approaches the far-field condition, i.e., $R_0$ and $R_1$ is sufficiently large, we have $b_1\approx0$. Under this condition, the optimal APV in \eqref{eqn_OptAPV_Lem1} can be approximated as
\begin{equation}
	\underset{b_1\rightarrow0}{\lim}x_n=\frac{\lambda\left(\frac{n}{N}+q_n\right)}{a_1}=\frac{\lambda\left(\frac{n}{N}+q_n\right)}{\cos\theta_1-\cos\theta_0},\forall n\ic{N},
\end{equation}
which is consistent with the results in \cite{zhu2023movable} derived under far-field propagation condition. This implies that our obtained results are more general and demonstrate the exceptional capability of MAs in both far- and near-field beam steering.
Next, we turn to the more general case of $K>1$, which is more challenging compared to the far-field case due to the nonlinear expressions of the phase difference in \eqref{eqn_Dist}. To tackle this challenge, we first consider a special case where the angle of arrivals for all users are identical, i.e., $\theta_k=\theta_0$, $\forall k$, and introduce the following two lemmas.

\begin{lemma}
	Let $N_1$ and $N_2$ be two positive integers, and define $N\triangleq N_1\times N_2$. Then, for any integer $n\in\left\{1,2,\cdots,N\right\}$, there exists a \textit{unique} pair $(n_1,n_2)\in[1,N_1]\times[1,N_2]$ such that $n$ can be represented by the tuple $(n_1,n_2)$.
\end{lemma}
\begin{IEEEproof}
	Note that we can define a mapping from each pair $(n_1,n_2)$ to $n$ via the following equation:
	\begin{equation}\label{eqn_Mapping}
		n=n_1+(n_2-1)\times N_1, \forall n=1,2,\cdots,N.
	\end{equation}
	Next, we check the uniqueness of this mapping. Suppose two different pairs $(n_1^{\prime},n_2^{\prime})$ and $(n_1,n_2)$ map to the same $n$. Then, according to \eqref{eqn_Mapping}, we have
	\begin{equation}\label{eqn_Lem2_Eq1}
		n_1^{\prime}+(n_2^{\prime}-1)\times N_1=n_1+(n_2-1)\times N_1,
	\end{equation}
leading to
	\begin{equation}\label{eqn_Lem2_Eq2}
		\left(n_2^{\prime}-n_2\right)N_1=n_1-n_1^{\prime}.
	\end{equation}
	Now, the left-hand side of \eqref{eqn_Lem2_Eq2} is the multiplier of $N_1$, while its right-hand side is bounded by $\pm\left(N_1-1\right)$. The only condition under which this equation holds is $n_1=n_1^{\prime}$ and $n_2=n_2^{\prime}$, which is contradictory to the presumption $(n_1^{\prime},n_2^{\prime})\ne(n_1,n_2)$. Hence, the mapping in \eqref{eqn_Mapping} is unique. This completes the proof.
\end{IEEEproof}
\begin{lemma}
    If an APV $\bs{x}\ib{R}^{N_1\times 1}$ satisfies \eqref{eqn_SV} for $N=N_1$ and $K=K_1$, i.e., $f_k(\bs{x})=0$, $k=1,2,\cdots,K_1$. Then, there exists an APV $\bs{x}^{\star}=\left[x^{\star}_1,x^{\star}_2,\cdots,x^{\star}_{N_1N_2}\right]^T\ib{R}^{N_1N_2\times 1}$ that satisfies the \eqref{eqn_SV} for $N=N_1N_2$ and $K=K_1+1$ for any $N_2\ge2$, i.e.,  $f_k(\bs{x}^{\star})=0$, $k=1,2,\cdots,K_1+1$.
\end{lemma}
\begin{IEEEproof}
	First, under the conditions $\theta_k=\theta_0$, $\forall k$, we have $a_k=0$, $\forall k$. According to Lemma 2, for any integer $n\in\left\{1,2,\cdots,N_1N_2\right\}$, it can be uniquely represented by the tuple $(n_1,n_2)\in[1,N_1]\times[1,N_2]$. Thus, the coordinate of the $n$-th MA, i.e., $x_n$, can be uniquely and equivalently represented by $x_{n_1+(n_2-1)N_1}$, and the beam gain at user $k$ in \eqref{eqn_SV} can be rewritten as
	\begin{equation}\label{eqn_Lem3_Eq1}
		f_k(\bs{x})=\sum_{n=1}^{N}e^{j\frac{2\pi}{\lambda}b_kx_n^2}=\sum_{n_1=1}^{N_1}\sum_{n_2=1}^{N_2}e^{j\frac{2\pi}{\lambda}b_kx_{n_1+(n_2-1)N_1}^2}.
	\end{equation}
	Next, we construct $x_{n_1+(n_2-1)N_1 }$ as
	\begin{equation}\label{eqn_Lem3_Eq2}
		x_{n_1+(n_2-1)N_1}=\sqrt{x_{n_1}^2+(n_2-1)d},\forall n_1,n_2,
	\end{equation}
	where $d$ is a constant. By substituting \eqref{eqn_Lem3_Eq2} into \eqref{eqn_Lem3_Eq1}, we have
		\begin{align}
f_k(\bs{x})&=\sum_{n_1=1}^{N_1}\sum_{n_2=1}^{N_2}e^{j\frac{2\pi}{\lambda}b_k\left[x_{n_1}^2+(n_2-1)d\right]} \nonumber\\
			&=\sum_{n_1=1}^{N_1}e^{j\frac{2\pi}{\lambda}b_kx_{n_1}^2}\times\sum_{n_2=1}^{N_2}e^{j\frac{2\pi}{\lambda}b_k(n_2-1)d},
		\end{align}
	where $k=1,2\cdots,K_1+1$. Then, for $1\le k \le K$, it is easy to verify $\sum_{n_1=1}^{N_1}e^{j\frac{2\pi}{\lambda}b_kx_{n_1}^2}=0$, and we have $f_k(\bs{x})=0$, $1\le k \le K$. In addition, for $k=K_1+1$, we set
	\begin{equation}
		d=\frac{(1/N_2+q)\lambda}{|b_{K_1+1}|},
	\end{equation}
	where $q$ is an integer that ensures the constraints in \eqref{eqn_Cons_MinDist}, leading to $f_{K_1+1}(\bs{x})=0$. Therefore, by setting $x_{n}$ according to \eqref{eqn_Lem3_Eq2}, the APV $\bs{x}^{\star}$ can achieve $f_k(\bs{x}^{\star})=0$, $k=1,2,\cdots,K_1+1$. This completes the proof.
\end{IEEEproof}
Based on these lemmas, we can construct an optimal APV under some specified values of $N$ and $K$. If $N$ can be decomposed into the product of several integers as $N=\prod_{i=1}^{I(N)}g_i$, where $I(N)$ represents the total number of factors of $N$ and all factors are sorted in an ascending order as $g_1\le g_2 \le \cdots \le g_{I(N)}$. Then, we can always construct an APV $\bs{x},$ such that for any given $K\le I(N)$, the orthogonality constraints in \eqref{eqn_SV} can be satisfied. The procedures of the construction are summarized in Algorithm 1.

\vspace{-9pt}
\begin{algorithm}[!t]
	\caption{Proposed Algorithm for Constructing an Optimal APV}
	\label{alg_SU}
	\begin{algorithmic}[1]
		\STATE Calculate the prime factorization of $N$ as $N\!=\!\prod_{i=1}^{I(N)}g_i$.
		\IF{$K<I(N)$}
			\STATE Set $g_{K}=\prod_{i=K}^{I(N)}g_i$.
			\STATE Set $I(N)=K$.
		\ENDIF
		\FOR{$i=1\rightarrow I(N)$}
			\IF{$i=1$}
				\STATE Calculate $x_n$, $n=1,2,\cdots,g_1$, based on Proposition 1.
			\ELSE
				\STATE Calculate $x_{n}$, $n=1,2,\cdots,g_i$, by setting $N_1=\prod_{j=1}^{i-1}g_j$, $N_2=g_i$, and applying Lemma 2.
			\ENDIF
		\ENDFOR
		\STATE Output the APV $\bs{x}$ as the solutions to (P2).
	\end{algorithmic}
\end{algorithm}

\subsection{Scenario 2: Multi-Beam Forming}
In this subsection, our main aim is to show that under the MRT for user 0, a full beam gain can also be achieved at the $K$ other users, i.e., 
\begin{align}\label{eqn_BeamGain_w_MRT}
   G\left(\bs{x},R_k,\theta_k\right)&=\frac{1}{N}\left|\bs{a}(\bs{x},R_0, \theta_0)^{H}\bs{a}(\bs{x},R_k, \theta_k)\right|^2 \nonumber\\
  &=\frac{1}{N}\left|\sum_{n=1}^{N}e^{j\frac{2\pi}{\lambda}\left(a_kx_n+b_kx_n^2\right)}\right|^2,\forall k\ic{K},
  \end{align}
by optimizing the APV. To this end, we present the following theorem.
\begin{theorem}
	Define $\hat{a}_k\triangleq a_k/\lambda$ and $\hat{b}_k=\sqrt{b_k/\lambda}$. If all $\hat{a}_k$ and $\hat{b}_k$ are rational numbers, there must exist an APV $\bs{x}^{\star}$ such that $G\left(\bs{x},R_k,\theta_k\right)=N$, $\forall k\ic{K}$.
\end{theorem}
\begin{IEEEproof}
	To construct the desired APV, we place the $N$ MAs along the transmit array with an equal spacing $d$. As such, the coordinate of the $n$-th MA is given by $x_n=(n-1)d$, $\forall n\ic{N}$. Then, the beam gain at the $k$-th user in \eqref{eqn_BeamGain_w_MRT} can be expressed as
	\begin{equation}\label{eqn_Th1_Eq1}
		G\left(\bs{x},R_k,\theta_k\right)=\frac{1}{N}\left|\sum_{n=1}^{N}e^{j2\pi\left[\left(n-1\right)\hat{a}_kd+\left(n-1\right)^2\left(\hat{b}_kd\right)^2\right]}\right|^2.
	\end{equation}
	To reap the maximum beam gain in \eqref{eqn_Th1_Eq1}, we can adjust the spacing $d$ such that each exponential term in \eqref{eqn_Th1_Eq1} is equal to unity, which is equivalent to
	\begin{equation}\label{eqn_Th1_Eq2}
		\hat{a}_kd=m_{k,1},\quad \hat{b}_kd=m_{k,2},\quad\forall k\ic{K}.
	\end{equation}
	where $m_{k,i}$, $i=1,2$, are integers. Note that \eqref{eqn_Th1_Eq2} is equivalent to find a set of $m_{k,i}$'s such that
		\begin{align}\label{eqn_Th1_Eq3}
			d&=\frac{m_{1,1}}{\hat{a}_1}=\frac{m_{2,1}}{\hat{a}_2}=\cdots=\frac{m_{K,1}}{\hat{a}_K} \nonumber\\
			&=\frac{m_{1,2}}{\hat{b}_1}=\frac{m_{2,2}}{\hat{b}_2}=\cdots=\frac{m_{K,2}}{\hat{b}_K}.
		\end{align}
	Next, we show how to construct the desired $m_{k,i}$'s satisfying \eqref{eqn_Th1_Eq3}. According to the basic number theory, a rational number can be expressed as the ratio of two relatively prime integers. As each $\hat{a}_k$ and $\hat{b}_k$ are rational numbers, they can be respectively represented as $\hat{a}_k=\frac{p_{k,1}}{q_{k,1}}$, $\hat{b}_k=\frac{p_{k,2}}{q_{k,2}}$, $k\ic{K}$, where $p_{k,1}$ (or $p_{k,2}$) and $q_{k,1}$ (or $q_{k,2}$) are two relatively prime integers. Define
	\begin{equation}
		\hat{m}_{k,i}\triangleq\frac{p_{k,i}}{q_{k,i}}\prod_{j=1}^{K}\prod_{i=1}^{2}q_{j,i},\forall j\ic{K},i=1,2,
	\end{equation}
	and $c_{\max}$ denote the greatest common factor of all $\hat{m}_{k,i}$'s. As such, we can set $m_{k,i}=\frac{\hat{m}_{k,i}}{c_{\max}}$, $\forall k\ic{K}^c$, $i=1,2$, and the resulting antenna spacing $d$ is
	\begin{equation}
		d^{\star}=\frac{\zeta\prod_{j=1}^{K}\prod_{i=1}^{2}q_{j,i}}{c_{\max}},
	\end{equation}
	where $\zeta$ is the minimum integer that ensures $d^{\star}\ge D_{\min}$. By setting $x_n^{\star}=(n-1)d^{\star}$, the maximum beam gain can be achieved at all $K$ users. This completes the proof.
\end{IEEEproof}
Theorem 1 indicates that by merely adjusting antenna spacing, it is possible to generate multiple grating lobes and achieve maximum beam gain for an arbitrary number of users. Notably, Theorem 1 is derived under the assumption that all $\hat{a}_k$'s and $\hat{b}_k$'s are rational numbers. For cases where these values are irrational, Theorem 1 may also approximately hold by appropriately truncating each $\hat{a}_k$ and $\hat{b}_k$ into rational numbers. 

\section{Proposed Solutions to (P2) and (P3)}\label{Sec4}
In this section, we focus on solving (P2) and (P3) in the general case. First, we start from (P2) in Scenario 1.

\subsection{Proposed Solution to (P2)}
For (P1), due to the complexity of the objective function in terms of $\boldsymbol{x}$ involving matrix projection, we adopt a discrete sampling strategy to circumvent this difficulty. {This strategy differs from conventional gradient-based methods in that it directly explores a discretized feasible set, avoiding the need for differentiable objective functions and constraints. Hence, it is more robust to non-convexities and discrete constraints (e.g., minimum antenna spacing), and is less prone to local optima, as will be shown in Section \ref{Secsix} via simulation.}

Specifically, we discretize the MA's movement range into $M\;(M\gg N)$ discrete sampling points, with the distance between any two adjacent sampling points given by $d=D_{\max}/M$. As such, the position of the $i$-th sampling point is given by $x_{{\text{s},}i}=iD_{\max}/M,\;i\in \mathcal M \triangleq\{0,1,\cdots,M\}$. Let $\mathcal{S} =\{x_{{\text{s},}i}| i\in \mathcal{M}\}$ denote the set of all sampling points. By this means, the optimal solution to (P2) can be obtained by enumerating all feasible combinations of the $N$ MAs based on the sampling points in $\cal S$. However, this results in a prohibitively high searching complexity, especially in the case of a large value of $M$ for higher resolution. To tackle this challenge, we propose to optimize the positions of all MAs in a sequential manner, thereby obtaining a high-quality suboptimal solution.

Specifically, we conduct $R$ rounds of sequential search, and in each round, we sequentially update the positions of the $N$ MAs with $N$ iterations. In the $n$-th iteration, we update the position of the $n$-th MA, i.e., $x_n$, while keeping the positions of all other $(N-1)$ MAs fixed. Let $x_j^{(r)}$ denote the updated position of the $j$-th MA in the $r$-th round, $1 \le j \le n-1$. Hence, the set of all feasible sampling points for optimizing $x_n$ in this round is given by
\begin{align}\label{SnUpdate}
\mathcal{S}^{(r)}_n = \{s | s \in &\mathcal{S}, |s - x_j^{(r)}| \geq D_{\min},  1 \le j \le n-1, \nonumber\\
&|s - x_i^{(r-1)}| \geq D_{\min},  n+1 \le i \le N\},
\end{align}
for $2 \le n \le N-1$. In addition, we set $\mathcal{S}^{(r)}_1=\{s | s \in \mathcal{S}, |s - x_j^{(r-1)}| \geq D_{\min},  2 \le j \le N\}$ and $\mathcal{S}^{(r)}_N=\{s | s \in \mathcal{S}, |s - x_j^{(r)}| \geq D_{\min},  1 \le j \le N-1\}$. Then, we can determine $x^{(r)}_n$ as
 \begin{equation}\label{XnUpdate}
       x^{(j)}_n = \arg\max_{s \in \mathcal{S}^{(r)}_n} G(\boldsymbol{\hat{x}}), \;{\text{s.t.}}\;s\in \mathcal{S}^{(r)}_n
 \end{equation}
where $\boldsymbol{\hat{x}} = [x_1^{(r)}, x_2^{(r)}, \ldots,x_{n-1}^{(r)}, s,x^{(r-1)}_{n+1},\ldots,x^{(r-1)}_{N}]^T.$ Next, we proceed to update the position of the $(n+1)$-th MA. {Note that the above sequential update process yields a non-decreasing objective value of \eqref{eqn_P2}. Since the optimal value of \eqref{eqn_P2} is upper bounded, our proposed algorithm is ensured to converge.} After updating the positions of all MAs in the $r$-th round, the algorithm can return to update the position of the first MA in the $(r+1)$-th round. The overall algorithm is summarized in Algorithm 2. {It can be shown that the complexity of Algorithm 2 is on the order of $\mathcal{O}(RNM)$, which is linear in both $N$ and $M$.}

\begin{algorithm}[!ht]\label{A2}
  \caption{Overall Algorithm for Solving (P2)}
  \begin{algorithmic}[1]
  \STATE Initialize: $r \leftarrow 0$, $n \leftarrow 1$, and $x_{n}^{(0)}, n \in \cal N$.
  \WHILE{$r < R$}  
  \WHILE{$n \leq N$}
      \STATE Determine $\mathcal{S}^{(r+1)}_n$ based on \eqref{SnUpdate}
      \STATE Obtain $x_{n}^{(r+1)}$ based on \eqref{XnUpdate}.
      \STATE Update $n \leftarrow n+1$.
  \ENDWHILE
  \STATE Update $r \leftarrow r+1$.  
  \ENDWHILE
  \STATE Output $x_n^{(r)}$ and output the ZF beamforming based on \eqref{eqn_ZF}.
\end{algorithmic}
\end{algorithm}\vspace{-9pt}

\subsection{Proposed Solution to (P3)}
In this subsection, we propose an AO algorithm to solve (P3). Specifically, we decouple (P3) into two subproblems for optimizing $\boldsymbol{x}$ and $\boldsymbol{w}$, respectively, and solve each of them individually.

\emph{1) Optimization of $\boldsymbol{w}$ with a given $\boldsymbol{x}$:}
First, we aim to optimize $\boldsymbol{w}$ in (P3) with a given APV $\boldsymbol{x}$. Since constraint \eqref{eqn_BeamGain_GL} is non-convex with respect to $\boldsymbol{w}$, we use the successive convex approximation (SCA) technique to tackle it. By applying the first-order Taylor expansion at a given local point $\boldsymbol{w}^t$ in the $t$-th SCA iteration, we can obtain a linear surrogate function $\bar{G}(\boldsymbol{w},\boldsymbol{x},\theta_k,R_k|\boldsymbol{w}^t)$, i.e.,
\begin{align}
& G(\boldsymbol{w},\boldsymbol{x},\theta_k,R_k) \notag\\
& \geq\bar{G}(\boldsymbol{w},\boldsymbol{x},\theta_k,R_k|\boldsymbol{w}^t) \notag\\
& \triangleq G(\boldsymbol{w}^t,\boldsymbol{x},\theta_k,R_k)+ 2\mathrm{Re}\{(\boldsymbol{w}^t)^H\boldsymbol{\Gamma}(\boldsymbol{x}, \theta_k,R_k)(\boldsymbol{w}-\boldsymbol{w}^t)\} \notag\\
& =2\mathrm{Re}\{(\boldsymbol{w}^t)^{H}\boldsymbol{\Gamma}(\boldsymbol{x}, \theta_k,R_k)\boldsymbol{w}\}-G(\boldsymbol{w}^t,\boldsymbol{x},\theta_k,R_k), 
\end{align}
where \mbox{$\boldsymbol{\Gamma}(\boldsymbol{x}, \theta_k,R_k)\triangleq\boldsymbol{\alpha}(\boldsymbol{x}, \theta_k,R_k) \boldsymbol{\alpha}(\boldsymbol{x}, \theta_k,R_k)^H\in\mathbb{C}^{N \times N}$}. 

By replacing $G(\boldsymbol{w},\boldsymbol{x},\theta_k,R_k)$ in constraint $\eqref{eqn_BeamGain_GL}$ with $\bar{G}(\boldsymbol{w},\boldsymbol{x},\theta_k,R_k|\boldsymbol{w}^t)$, (P3) can be recast as
\begin{subequations}\label{eqn_OptPrblm_P4}
  \begin{align}
    {\text{(P3.$t$)}}\quad &\underset{\boldsymbol{w},\delta} {\max}\quad  \delta
    \\
    & \bar{G}(\boldsymbol{w},\boldsymbol{x},\theta_k,R_k|\boldsymbol{w}^t)\geq\delta, \ k=0,1,\ldots,K, \label{eqn_RelaxGain_GL}\\
    & \|\boldsymbol{w}\|_{2}\leq1.\label{eqn_ConsP4_Power}
    \end{align}
    \end{subequations}
in the $t$-th SCA iteration. Note that (P3.$t$) is a convex optimization problem, which can be optimally solved via the interior-point algorithm. Then, we proceed to solve (P3.$t+1$), where the local point $\boldsymbol{w}^t$ is updated as the optimal solution to (P3.$t$), until convergence is achieved.

\emph{2) Optimization of $\boldsymbol{x}$ with a given $\boldsymbol{w}$:}
Second, we optimize the APV $\boldsymbol{x}$ in problem (P3) with a given $\boldsymbol{w}$. Given $\boldsymbol{w}$, problem (P3) can be expressed as
\begin{subequations}\label{eqn_OptPrblm_P2}
  \begin{align}
    {\text{(P4)}}\quad &\underset{\boldsymbol{x},\delta_{\boldsymbol{x}}} {\max}\quad  \delta_{\boldsymbol{x}}
    \\
    \mathrm{s.t.}\quad  & G(\boldsymbol{x}, \theta_k, R_k)\geq\delta_{\boldsymbol{x}}, \ k=0,1,\ldots,K, \\
    &\eqref{eqn_Cons_Region},\eqref{eqn_Cons_MinDist}.\nonumber
    \end{align}
    \end{subequations}
Notably, (P4) has a similar form to (P2) except the objective function. Therefore, we can still use the sequential update method as in solving (P2). The details are omitted for brevity. Let $\boldsymbol{x}^\star$ denote the optimized solution to (P4).
{Notably, the convergence of the proposed AO algorithm is guaranteed. This is because its two subproblems are solved using iterative algorithms that each guarantee a monotonic increase in the objective value of (P3). Since the optimal value of (P3) is upper-bounded, the overall procedure is therefore guaranteed to converge.}

\emph{3) Overall Algorithm:}
The details of the proposed AO algorithm are summarized in Algorithm 3. The complexity of Algorithm 3 arises mainly from solving (P3.$t$) and (P4). {The former yields a complexity order of $\mathcal{O}(K^{0.5}N(N^2+K))$\cite{wang2014outage}, while the latter yields a complexity order of $\mathcal{O}(NMI_1)$, where $I_1$ denotes the total rounds of sequential updates involved}.
\begin{algorithm}[!ht]
  \renewcommand{\algorithmicrequire}{\textbf{Input:}}
	\renewcommand{\algorithmicensure}{\textbf{Output:}}
  \caption{Overall Algorithm for Solving (P3)}
  \begin{algorithmic}[1]
  \STATE Initialize: $t \leftarrow 0 , i \leftarrow 0, \boldsymbol{w}^0$.
  \WHILE{AO convergence is not achieved}
  \WHILE{SCA convergence is not achieved}
      \STATE Solve problem (P3.$t$) with $\boldsymbol{x}=\boldsymbol{x}(i)$ using the interior-point algorithm to obtain $\boldsymbol{w}^{t+1}$.
      \STATE $t \leftarrow t+1$.
  \ENDWHILE
  \STATE Update $\boldsymbol{w}(i+1) \leftarrow \boldsymbol{w}^{t}, t \leftarrow 0$.
  \STATE Solve problem (P4) with $\boldsymbol{w}=\boldsymbol{w}(i+1)$ via the sequential update method.
  \STATE Update $\boldsymbol{x}(i+1) \leftarrow \boldsymbol{x}^\star$ and $i \leftarrow i+1$.
  \ENDWHILE
    \STATE \textbf{return} $\boldsymbol{x}(i),\boldsymbol{w}(i)$.
    \end{algorithmic}
\end{algorithm}

\section{Antenna Position Errors Analysis}
In this section, we present performance analysis regarding the impact of antenna position errors on the beamforming performance in both scenarios. As discussed in Section \ref{sc-bn}, it is possible to achieve beam nulling or grating lobes at the other $K$ targets with the MRT toward user 0. To facilitate our analysis, we assume this MRT in this section, i.e., $\bs{w}= \bs{a}(\bs{x},R_0, \theta_0)/\sqrt{N}$. Then, the beam gain at user $k, \forall k \in {\cal K}$ is given by
\begin{equation}\label{eq_mrt_gain}
G(\bs{x}, R_k, \theta_k) = \frac{1}{N}\left|\sum^N_{n=1}\exp\left[j\frac{2\pi}{\lambda}(r_{0,n} - r_{k,n})\right]\right|^2,
\end{equation}
where $r_{k,n}$ is defined in \eqref{eqn_Dist}.

Denote the position error for the $n$-th MA as $\Delta d_n$; therefore, its actual position becomes $x_n + \Delta d_n$. In this paper, we assume that the maximum position deviation is upper-bounded by a given constant $\epsilon$, i.e., $\lvert \Delta d_n \rvert \le \epsilon$, $\forall n \in \cal{N}$. Let ${\bs{\Delta d}}=[\Delta d_1, \Delta d_2, \cdots, \Delta d_N]^T$ denote the vector of antenna position errors. Then, the actual beam gain at user $k$ is given by $G({\bs x}^*+\Delta\bs{d}, R_k, \theta_k)$. Next, we characterize its worst-case performance in the two scenarios considered, respectively.

In Scenario 1, let ${\bs x}^*=[x_1^*,x_2^*,\cdots,x_N^*]^T$ denote an APV solution that satisfies $G(\bs{x}, R_k, \theta_k)=0, \forall k \in \cal K$, which is calculated assuming that there are no position errors. In the presence of the position error ${\bs{\Delta d}}$, we aim to derive its worst-case impact on the beam nulling performance. To this end, we consider the sum of the beam gain at the $K$ undesired users as a performance metric, which is desired to be zero. As such, the worst-case performance can be characterized by solving the following optimization problem,
\begin{subequations}\label{eq_problem_zf}
\begin{align}
\mathop{\mathrm{max}}_{\{\Delta d_n\}} ~&\sum_{k=1}^K G({\bs x}^*+\Delta\bs{d}, R_k, \theta_k) \label{eq_problem_zf_a}\\
\mathrm{s.t.}~~&|\Delta d_n| \leq \epsilon, \forall n \in \cal{N}. \label{eq_problem_zf_b}
\end{align}
\end{subequations}
It is worth noting that in the case of $\epsilon=0$, the objective function of problem \eqref{eq_problem_zf} can achieve the minimum value of zero.

In Scenario 2, let $\tilde{\bs x}^*$ denote an APV solution that satisfies $G(\bs{x}, R_k, \theta_k)=N, \forall k \in \cal K$ in the absence of the position errors. Unlike Scenario 1, given $\tilde{\bs x}^*$ and the position error ${\bs{\Delta d}}$, we aim to derive the minimum (instead of maximum) sum of the beam gains towards the $K$ users, i.e.,
\begin{subequations}\label{eq_problem_gl}
\begin{align}
\mathop{\mathrm{min}}_{\{\Delta d_n\}} ~&\sum_{k=1}^K G(\tilde{\bs x}^* +\Delta\bs{d}, R_k, \theta_k) \label{eq_problem_gl_a}\\
\mathrm{s.t.}~~&|\Delta d_n| \leq \epsilon, \forall n \in \mathcal{N}. \label{eq_problem_gl_b}
\end{align}
\end{subequations}

Note that in the case with $\epsilon=0$, the objective value of problem \eqref{eq_problem_gl} can achieve the maximum value of $NK$. However, both optimization problems \eqref{eq_problem_zf} and \eqref{eq_problem_gl} are non-convex and difficult to tackle due to the complex exponential terms in the expressions of beam gains. In the following sections, we apply Taylor expansion to circumvent this difficulty. For notational simplicity, we express the optimal APV solutions in both scenarios (i.e., $x_n^*$ and ${\tilde x}_n^*)$ as $x_n$, $n \in {\cal N}$, without highlighting their optimality, in the sequel of this section.\vspace{-6pt}

\subsection{Worst-Case Performance in Scenario 1}
To solve problem \eqref{eq_problem_zf}, we first apply Taylor expansion to linearize the near-field distance expression. Specifically, for the user at $(R_k, \theta_k)$, its distance with the $n$-th MA given the position error $\Delta d_n$ can be obtained by replacing $x_n$ in \eqref{eqn_Dist} with $x_n+\Delta d_n$, i.e., $r_{k, n}(x_n+\Delta d_n), n \in \cal{N}$. Next, we approximate it with the first-order Taylor expansion around $\Delta d_n = 0$, i.e.,
\begin{equation}\label{eq_distance_taylor_re}
r_{k,n}(x_n + \Delta d_n) \approx r_{k,n}^{(0)} + \beta_{k,n}\Delta d_n,
\end{equation}
where $r_{k,n}^{(0)} \triangleq r_{k,n}(x_n) = R_k - x_n \cos\theta_k + \frac{(x_n)^2}{2R_k}\sin^2 \theta_k$ is the nominal near-field distance without position error, and
\begin{equation}\label{eq_beta_re}
\beta_{k,n} = -\cos\theta_k + \frac{x_n}{R_k}\sin^2\theta_k,
\end{equation}
which captures the sensitivity of the distance w.r.t the position error.

Then, the phase difference between the desired user (user 0) and user $k$ at the $n$-th MA can be expressed as
\begin{align}
\Phi_{k,n}(x_n+\Delta d_n) &= \frac{2\pi}{\lambda}(r_{0,n}(x_n+\Delta d_n) - r_{k,n}(x_n+\Delta d_n)) \nonumber\\
&\approx \Phi_{k,n}^{(0)} + \Delta\Phi_{k,n},\label{eq_phase_diff_re}
\end{align}
where $\Phi_{k,n}^{(0)} = \frac{2\pi}{\lambda}\left(r_{0,n}^{(0)} - r_{k,n}^{(0)}\right)$ is the nominal phase difference without position errors, and $\Delta\Phi_{k,n} = \frac{2\pi}{\lambda}(\beta_{0,n}-\beta_{k,n})\Delta d_n$ represents the phase perturbation due to the position error.

Based on \eqref{eq_phase_diff_re}, we have
\[\exp[j\Phi_{k,n}(x_n+\Delta d_n)] \approx \exp[j(\Phi_{k,n}^{(0)} + \Delta\Phi_{k,n})].\] 
Let $s_{k,n} = \exp[j \Phi_{k,n}^{(0)}] = \exp\left[j \frac{2\pi}{\lambda} (r_{0,n}^{(0)} - r_{k,n}^{(0)})\right]$. Since $\Delta\Phi_{k,n}$ is small (due to the small position error $\Delta d_n$), we apply the first-order Taylor expansion to $\exp[j\Delta\Phi_{k,n}]$ around $\Delta\Phi_{k,n}=0$, leading to $\exp[j\Delta\Phi_{k,n}] \approx 1 + j\Delta\Phi_{k,n}$.
Hence, we have
\begin{equation}\label{eq_exp_taylor_re}
\exp[j(\Phi_{k,n}^{(0)} + \Delta\Phi_{k,n})] \approx s_{k,n}(1 + j\Delta\Phi_{k,n}).
\end{equation}

{Note that the accuracy of the first-order Taylor expansion $e^{j\Delta\phi} \approx 1 + j\Delta\phi$ is mathematically bounded by the remainder of the Taylor series, which is proportional to $0.5(\Delta\phi)^2$. In our context, modern stepper motors can achieve positioning accuracy of approximately $1/50$ of a wavelength (i.e., $\epsilon \le 0.02\lambda$). At this error level, the resulting phase perturbation $\Delta\Phi_{k,n}$ is sufficiently small to ensure that the higher-order error terms are negligible (typically far smaller than the first-order term), thereby guaranteeing the accuracy of the approximation.}

Define $AF(R_k, \theta_k)= {{\bs{w}}_{\bs{x}}^H}{\bs{a}}(\bs{x},R_k,\theta_k)$ as the array factor for user $k$, which is given by
\begin{equation}\label{eq_af_definition}
AF(R_k, \theta_k) = \frac{1}{\sqrt{N}} \sum_{n=1}^N \exp[j\Phi_{k,n}(x_n+\Delta d_n)].
\end{equation}

Based on \eqref{eq_exp_taylor_re}, the array factor can be approximated as
\begin{equation}\label{eq_af_approx}
AF(R_k, \theta_k) \approx \frac{1}{\sqrt{N}}\sum_{n=1}^N s_{k,n}(1 + j\Delta\Phi_{k,n}).
\end{equation}

Let $S_k^{(0)} \triangleq \sum_{n=1}^N s_{k,n}$ and $S_k^{(1)} \triangleq j \sum_{n=1}^N s_{k,n}\Delta\Phi_{k,n}$. Then, \eqref{eq_af_approx} can be rewritten as $AF(R_k, \theta_k) \approx (S_k^{(0)} + S_k^{(1))})/{\sqrt{N}}$. Thus the beam gain towards user $k$ can be expressed as
\begin{equation}\label{eq_gain_approx_re}
G(\bs{x}^* +\Delta\bs{d}, R_k, \theta_k) = |AF(R_k, \theta_k)|^2 \approx \frac{1}{N}\left|S_k^{(0)} + S_k^{(1)}\right|^2.
\end{equation}

For convenience, in the sequel of this paper, we omit the constant $1/N$ in \eqref{eq_gain_approx_re}. Recall that under the optimal antenna position vector $\bs{x}^*$, the interference toward each undesired user $k, k \in {\cal K}$ is nulled if $\Delta \bs{d}=\bs{0}$, i.e.,
\begin{equation}\label{eq_perfect_null}
G(\bs{x}^*, R_k, \theta_k) = |S_k^{(0)}|^2 = \left|\sum_{n=1}^N s_{k,n}\right|^2 = 0,
\end{equation}
which leads to $S_k^{(0)} \triangleq \sum_{n=1}^N s_{k,n} = 0$, $\forall k \in \cal{K}$.

Given that $S_k^{(0)} = 0$, the beam gain in \eqref{eq_gain_approx_re} is simplified as
\begin{equation}\label{eq_af_simplified}
G(\bs{x}^* +\Delta\bs{d}, R_k, \theta_k) = \lvert S_k^{(1)} \rvert^2 = \lvert \sum_{n=1}^N s_{k,n}\Delta\Phi_{k,n}\rvert^2.
\end{equation}

Since $\Delta{\Phi}_{k,n}$ = $\frac{2\pi}{\lambda}(\beta_{0,n}-\beta_{k,n})\Delta d_n$, we have
\begin{equation}\label{eq_zf_quadratic_term}
\begin{aligned}
G(\bs{x}^* + \Delta\bs{d}, R_k, \theta_k) = \Big(\frac{2\pi}{\lambda}\Big)^2\Bigg|\sum_{n=1}^N s_{k,n}(\beta_{0,n} - \beta_{k,n})\Delta d_n\Bigg|^2.
\end{aligned}
\end{equation}

For convenience, let $c_{k,n} = \frac{2\pi}{\lambda}s_{k,n}(\beta_{0,n} - \beta_{k,n})$. Then, the beam gain in \eqref{eq_zf_quadratic_term} can be rewritten as
\begin{equation}\label{eq_squared_magnitude_expansion}
\begin{split}
G(\bs{x}^* + \Delta\bs{d}, R_k, \theta_k) &= \left|\sum_{n=1}^N c_{k,n}\Delta d_n\right|^2,
\end{split}
\end{equation}
and the optimization problem in \eqref{eq_problem_zf} can be approximated as
\begin{subequations}\label{eq_problem_zf_approx}
\begin{align}
\mathop{\mathrm{max}}_{\Delta\bs{d}} ~&\sum_{k=1}^K \left|\sum_{n=1}^N c_{k,n}\Delta d_n\right|^2 \label{eq_problem_zf_approx_obj} \\
\mathrm{s.t.}~~&|\Delta d_n| \leq \epsilon, \forall n \in \mathcal{N}. \label{eq_problem_zf_approx_constr}
\end{align}
\end{subequations}

Let $\bs{c}_k = [c_{k,1}, c_{k,2}, \ldots, c_{k,N}]^T \in \mathbb{C}^{N\times1}$. Then, the objective function of problem \eqref{eq_problem_zf_approx} can be rewritten as
\begin{align}
\sum_{k=1}^K \left|\sum_{n=1}^N c_{k,n}\Delta d_n\right|^2
&= \Delta\bs{d}^T \left(\sum_{k=1}^K \bs{c}_k\bs{c}_k^H\right) \Delta\bs{d} .\label{eq_zf_obj_rewritten_matrix_form}
\end{align}

Let $\bs{Q} = \sum_{k=1}^K \bs{c}_k\bs{c}_k^H \in \mathbb{C}^{N \times N}$. Then, problem \eqref{eq_problem_zf_approx} becomes
\begin{subequations}\label{eq_problem_q_matrix}
\begin{align}
\max_{\Delta\bs{d}} \quad & \Delta\bs{d}^T \bs{Q} \Delta\bs{d} \label{eq_problem_q_matrix_obj}\\
\text{s.t.} \quad & |\Delta d_n| \leq \epsilon, \quad \forall n \in \{1, 2, \dots, N\}. \label{eq_problem_q_matrix_constr}
\end{align}
\end{subequations}

Note that problem \eqref{eq_problem_q_matrix} is a non-convex QCQP problem, as the objective function is convex (rather than concave) in terms of $\bs{\Delta d}$. To tackle this challenge, we adopt the SDR technique by letting $\bs{X} = \Delta\bs{d} \Delta\bs{d}^T$, with ${\text{rank}(\bs{X})} = 1$. With this transformation, the objective function becomes
\begin{equation}\label{eq_obj_trace_form}
\Delta\bs{d}^T \bs{Q} \Delta\bs{d} = \text{Tr}(\Delta\bs{d}^T \bs{Q} \Delta\bs{d}) = \text{Tr}(\bs{Q}\bs{X}).
\end{equation}

Moreover, the constraints $|\Delta d_n| \leq \epsilon, n\in \cal{N}$, can be transformed into the following constraints on the diagonal elements of $\bs{X}$, i.e., 
\begin{equation}\label{eq_X_diag_constraint_SDR}
{\bs e}^T_n {\bs X} {\bs e}_n = (\Delta d_n)^2 \leq \epsilon^2, \quad \forall n \in \cal{N},
\end{equation}
where $\bs{e}_n$ denotes the $n$-th column of the identity matrix $\bs{I}_N$.

By relaxing the constraint $\text{rank}(\bs{X}) = 1$, problem \eqref{eq_problem_q_matrix} can be transformed into the following semidefinite programming (SDP) problem,
\begin{subequations}\label{eq_problem_trace}
\begin{align}
\max_{\bs{X}} \quad & \text{Tr}(\bs{Q}\bs{X}) \label{eq_problem_trace_obj}\\
\text{s.t.} \quad & \bs{X} \succeq 0 \label{eq_problem_trace_constr_psd}\\
& {\bs e}^T_n {\bs X} {\bs e}_n \leq \epsilon^2, \quad \forall n \in \cal{N}. \label{eq_problem_trace_constr_diag}
\end{align}
\end{subequations}
Problem \eqref{eq_problem_trace} can be optimally solved using interior-point methods with computational complexity order of $\mathcal{O}(N^{3.5})$\cite{luo2010semidefinite}. Denote by $\bs{X}^*$ the optimal solution to the relaxed problem \eqref{eq_problem_trace}. Next, we need to recover a feasible position error vector $\Delta\bs{d}^*$ from it.

Evidently, if $\text{rank}(\bs{X}^*)=1$, we can directly obtain $\Delta\bs{d}^*$ from $\bs{X}^*$ via eigenvalue decomposition. However, if $\text{rank}(\bs{X}^*) > 1$, we have to employ an approximation technique to obtain a rank-one solution. To this end, we adopt the well-known Gaussian randomization method: we compute the eigen-decomposition of $\bs{X}^*$ as $\bs{X}^* = {\bs V}^{T}\bs{\Lambda}\bs{V}$, where $\bs{V}$ is an orthogonal matrix whose columns are the eigenvectors of $\bs{X}^*$, and $\bs{\Lambda}$ is a diagonal matrix whose entries are the corresponding eigenvalues. Then, we can generate $\Delta\tilde{\bs{d}}$ as $\Delta \bs{d} = \bs{V}\bs{\Lambda}^{1/2}\bs{z}$, where $\bs{z} \sim \mathcal{N}({\bs{0}, \bs{I}})$ is a Gaussian distributed random vector. Given a sufficient number of realizations of $\bs{z}$, we select the one yielding the maximum objective value of problem \eqref{eq_problem_q_matrix} as the optimized solution.\vspace{-6pt}

\subsection{Worst-Case Performance in Scenario 2}
To solve \eqref{eq_problem_gl}, the beam gain at each user $k$ can be approximated by employing a similar first-order Taylor expansion used in Scenario 1, as detailed in \eqref{eq_phase_diff_re}-\eqref{eq_af_simplified}. In the case of multi-beam forming, it can be shown that the beam gain at $(R_k, \theta_k)$ can be approximated as
\begin{equation}\label{eq_gl_bg}
G(\tilde{\bs{x}}^* + \Delta \bs{d}, R_k, \theta_k) \approx  N + \sum_{n=1}^N \eta_{k,n} \Delta d_n,
\end{equation}
where
\begin{equation}\label{eq_eta_kn_definition}
\eta_{k,n} = -2\Re\left\{\frac{2\pi}{\lambda}j(\beta_{0,n} - \beta_{k,n})e^{j\Phi_{k,n}^{(0)}}\right\},
\end{equation}
with $\beta_{0,n}$ and $\beta_{k,n}$ already defined in \eqref{eq_beta_re}, and ${\bf{\Phi}}_{k,n}^{(0)}$ defined following \eqref{eq_phase_diff_re}. The details of deriving \eqref{eq_gl_bg} are omitted for simplicity.

Then, problem \eqref{eq_problem_gl} can be approximated as
\begin{subequations}\label{eq_grating_linear_re}
\begin{align}
\min_{\{\Delta d_n\}} \quad & \sum_{n=1}^N D_n\Delta d_n \label{eq_grating_linear_re_obj} \\
\mathrm{s.t.} \quad & |\Delta d_n| \leq \epsilon, \quad \forall n \in \mathcal{N}.\label{eq_grating_linear_re_constraint}
\end{align}
\end{subequations}
where $D_n = \sum_{k=1}^K \eta_{k,n}$ captures the aggregate sensitivity of the $n$-th antenna position error for all $K$ users.

For problem \eqref{eq_grating_linear_re}, it is not difficult to see that the optimal position errors are given by
\begin{equation}\label{eq_best_case_deriv_re}
\Delta d_n^\star = -\epsilon\,\mathrm{sign}(D_n), \quad n \in \mathcal{N}.
\end{equation}
Substituting \eqref{eq_best_case_deriv_re} into \eqref{eq_gl_bg}, the beam gain at user $k$ can be rewritten as
\begin{equation}\label{eq_best_case_gain_k_re}
G(\tilde{\bs{x}}^*+\Delta\bs{d}^\star, R_k, \theta_k) \approx N - \epsilon \sum_{n=1}^N \eta_{k,n} \,\mathrm{sign}(D_n).
\end{equation}
The worst-case sum beam gain over all users in ${\cal K}$ is thus given by
\begin{equation}\label{eq_total_best_case_gain_re}
\sum_{k=1}^K G(\tilde{\bs{x}}^*+\Delta\bs{d}^\star, R_k, \theta_k) \approx KN - \epsilon \sum_{n=1}^N |D_n|.
\end{equation}

{\it Theoretical Analysis of \eqref{eq_total_best_case_gain_re}:} It follows from \eqref{eq_total_best_case_gain_re} that the worst-case sum beam gain deviates from the ideal gain \( KN \) by an amount of \( \epsilon \sum_{n=1}^N |D_n|\). This performance gap is linearly proportional to both the maximum error bound $\epsilon$ and the sum-magnitude of the sensitivity factors, $|D_n|$. The term $D_n$, defined as the sum of individual user sensitivities $\eta_{k,n}$ (i.e., $D_n = \sum_{k=1}^K \eta_{k,n}$), aggregates the impact of the $n$-th antenna's position error for all $K$ users. 

More specifically, the sensitivity factor $\eta_{k,n}$, as defined in \eqref{eq_eta_kn_definition}, contains the term $2\pi/\lambda$. Consequently, $|D_n|$ is directly proportional to the operating frequency ($f = c/\lambda$). This implies that a higher frequency (or smaller wavelength $\lambda$) leads to a more pronounced performance degradation relative to the ideal sum gain $KN$. {This is expected, as a higher frequency results in a more rapid phase variation for the same physical displacement, making the system more sensitive to position errors.}

{Furthermore, the user locations, characterized by both distance $R_k$ and angle $\theta_k$, can also affect the performance gap through the term $\beta_{0,n} - \beta_{k,n}$ within $\eta_{k,n}$ in \eqref{eq_eta_kn_definition}. Based on the definition of $\beta_{k,n}$ in \eqref{eq_beta_re}, the impact of user geometry is reflected in two primary ways. {\it First}, regarding the angular impact, the error sensitivity is governed by the difference $| \beta_{0,n} - \beta_{k,n} |$, which is dominated by the term $| \cos\theta_0 - \cos\theta_k |$. This indicates that the performance degradation depends on the cosine difference between user angles. Consequently, for a fixed angular separation, the error sensitivity is maximized when users are located near the broadside direction (i.e., $\theta \approx \pi/2$), where the gradient of the cosine function is steepest. Moreover, a larger angular separation between user 0 and user $k$ generally leads to a larger magnitude for this difference, thereby exacerbating the performance degradation. This theoretical finding will be validated in Section VI-C via simulation, where we show that increasing the angular separation significantly amplifies the impact of position errors.}

{{\it Second}, regarding the distance impact, the term $\frac{x_n}{R_k}\sin^2\theta_k$ in \eqref{eq_beta_re} captures the near-field wavefront curvature. As user $k$'s distance $R_k$ decreases, the magnitude of this term becomes more significant, thus increasing the values of $\beta_{k,n}$ and $|D_n|$. This implies that the sensitivity of the phase error to position deviations scales approximately with $1/R_k$. To compare the sensitivity in near-field and far-field scenarios, we can let $R_k \rightarrow \infty$. In this case, the curvature term $\frac{x_n}{R_k}\sin^2\theta_k$ vanishes, and $\beta_{k,n}$ reduces to a constant $-\cos \theta_k$, which corresponds to the far-field planar wavefront model. This indicates that the near-field scenario introduces an additional sensitivity term due to curvature, making it inherently more susceptible to position errors than the far-field scenario.}

{In summary, the derived closed-form expression in \eqref{eq_total_best_case_gain_re} reveals that the resilience of MA-enhanced multi-beam forming to position errors diminishes with higher operating frequencies, stronger near-field effects (smaller $R_k$), and wider angular spreads among users.}\vspace{-6pt}

\section{Numerical Results}\label{Secsix}
In this section, numerical results are provided to validate our performance analysis and evaluate the effectiveness of our proposed algorithms. Unless otherwise stated, the simulation parameters are set as follows. The carrier wavelength is $\lambda=0.06$ meter (m), and the numbers of antennas is $N=6$. The minimum distance between any two MAs is set to $D_{\min}=\lambda/2$, and the size of the antenna array is $D_{\max}=9\lambda=0.54$ m. {With a maximum aperture $D_{\max} = 9\lambda$ at 5 GHz ($\lambda = 0.06$ m), the Rayleigh distance is $2D_{\max}^2/\lambda \approx 9.72$ m, corresponding to typical indoor communication ranges.}  {In the simulation, the users are uniformly distributed within the angular range $[0, \pi]$ and distance range $[3, 9.7]$ meters from the BS.} All results in this section are averaged over $10^3$ random generations of the users' positions. Moreover, we consider the following four benchmark schemes for performance comparison:
\begin{enumerate}
    \item {\textbf{Far-Field (FF)}: To highlight the advantages of near-field beamforming, we compare our proposed near-field design with a far-field design (assuming planar wavefronts). In this benchmark, $\bs{x}$ and $\bs{w}$ are optimized under far-field steering vectors.}
	\item \textbf{Particle swarm optimization (PSO)}: The APVs in Algorithms 2 and 3 are optimized in a continuous space via the PSO algorithm, following a similar approach to \cite{xiao2024multiuser}, while the transmit beamforming in Scenario 2 is optimized via the SCA algorithm.
	\item \textbf{Antenna selection (AS)}: In this benchmark, $D_{\max}/D_{\min}$ FPAs are deployed within the antenna array and separated by the minimum distance $D_{\min}$. Among them, $N$ antennas are selected for transmission. The associated optimization problem can be solved by Algorithms 2 and 3 by setting $d$ = $D_{\min}$.
	\item \textbf{Sparse array (SA)}: The $N$ antennas are deployed with an equal spacing of $D_{\max}/N$ along the antenna array. The transmit beamforming in Scenario 2 is optimized via the SCA algorithm.
	\item \textbf{FPA}: The $N$ antennas are deployed symmetrically to the center of the array and separated by the minimum distance $D_{\min}$. The transmit beamforming in Scenario 2 is optimized via the SCA algorithm.
\end{enumerate}\vspace{-6pt}

\subsection{Scenario 1: Beam Nulling}
\begin{figure}[!t]
	\centering
	\captionsetup{justification=raggedright,singlelinecheck=false}
	\includegraphics[width=7.6cm]{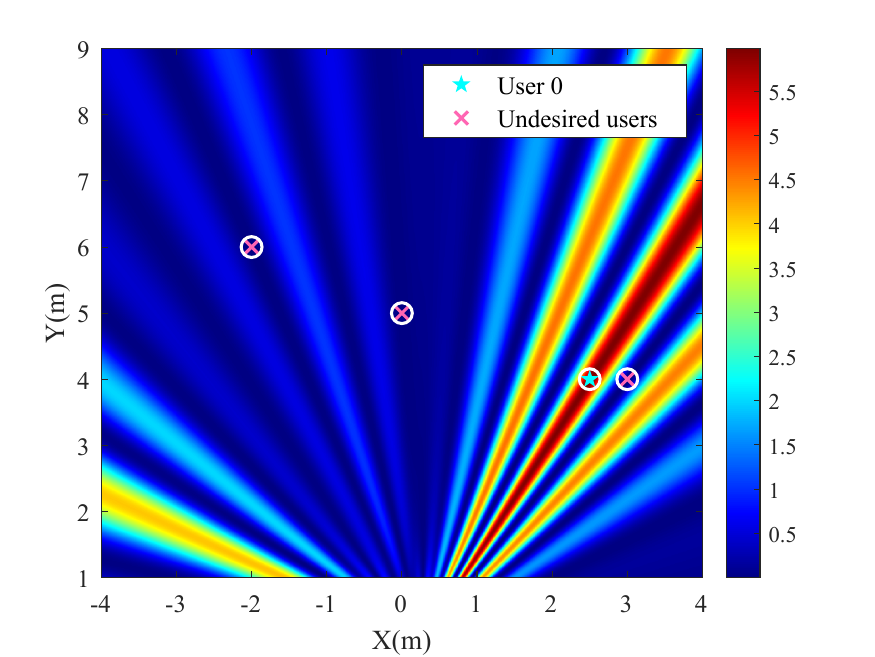}
	\caption{Distribution of beam gain in Scenario 1.}
	\label{Fig_S1_Heatmap}
    \vspace{-6pt}
\end{figure}
First, we consider the beam-nulling scenario and plot in Fig.~\ref{Fig_S1_Heatmap} the distribution of beam gain by Algorithm 2. The target user (i.e., user 0, marked by a blue pentagram) is positioned at $(R_0, \theta_0) = (4.72\;{\text{m}}, 1.01\;{\text{rad}})$. We consider $K=3$ undesired users (marked by red crosses) located at $(R_1, \theta_1) = (6.32\;{\text{m}}, 1.89\;{\text{rad}})$, $(R_2, \theta_2) = (5\;{\text{m}}, 1.57\;{\text{rad}})$, and $(R_3, \theta_3) = (5\;{\text{m}}, 0.93\;{\text{rad}})$, respectively. It is observed from Fig.~\ref{Fig_S1_Heatmap} that by optimizing the APV, the antenna array can generate a high beam gain (approximately 99.61\% of the full beam gain) at the target user. Meanwhile, the beam gains at all undesired users are effectively nulled. Furthermore, it can be seen that the high-gain region is concentrated around the vicinity of the target user, instead of forming an extended linear high-gain region typically observed under far-field propagation conditions. This observation indicates that our proposed algorithm can achieve beam focusing and nulling at the same time for the considered user positions.

\begin{figure}[!t]
	\centering
	\captionsetup{justification=raggedright,singlelinecheck=false}
	\includegraphics[width=7.6cm]{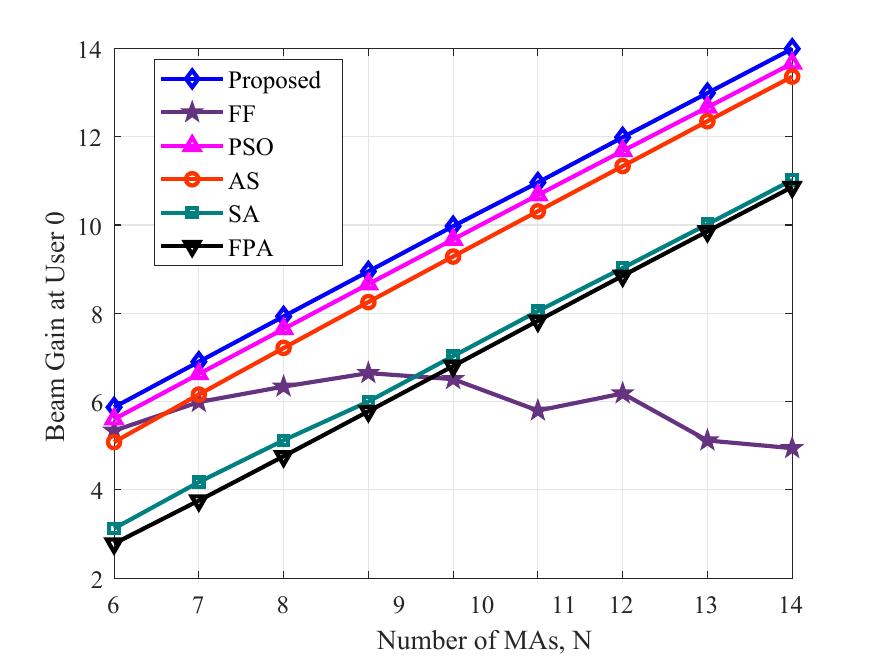}
	\caption{Beam gain at user 0 versus number of MAs in Scenario 1.}
	\label{Fig_S1_TxNums}
    \vspace{-9pt}
\end{figure}
Next, we plot in Fig.~\ref{Fig_S1_TxNums} the beam gain at user 0 versus the number of antennas, i.e., $N$, in Scenario 1, with the number of undesired users and the size of the movement region set to $K = 3$ and $D_{\max} = 1.5N\lambda$, respectively. It is observed that the performance of all considered schemes improves with the number of MAs. Specifically, the proposed algorithm achieves $99.34\%$ of the full beam gain (i.e., $N$) , on average and a $153\%$ improvement compared to the FPA benchmark. {The far-field benchmark exhibits limited performance improvement as the number of antennas $N$ increases, attaining only about $35.71\%$ of the gain achieved by the proposed design at $N=14$. This is because under near-field conditions, the spherical wavefront induces significant phase variations that cannot be accurately captured by the planar-wave assumption inherent in far-field models. Consequently, the far-field beamforming fails to precisely focus energy toward the target user subject to the beam nulling condition.} In addition, there exists an obvious performance gap between the proposed algorithm and the AS benchmark thanks to the larger degrees of freedom exploited by the former. Moreover, the proposed algorithm requires much fewer antennas to achieve the same beam gain compared to the FPA benchmark. For example, to achieve the beam gain of 6 (about 7.8 dB), the proposed algorithm needs only 6 antennas, whereas the FPA benchmark needs around 9 antennas. Moreover, the PSO benchmark is observed to achieve relatively worse performance compared to the proposed algorithm with higher computational complexity.

\begin{figure}[!t]
	\centering	\captionsetup{justification=raggedright,singlelinecheck=false}
	\includegraphics[width=7.6cm]{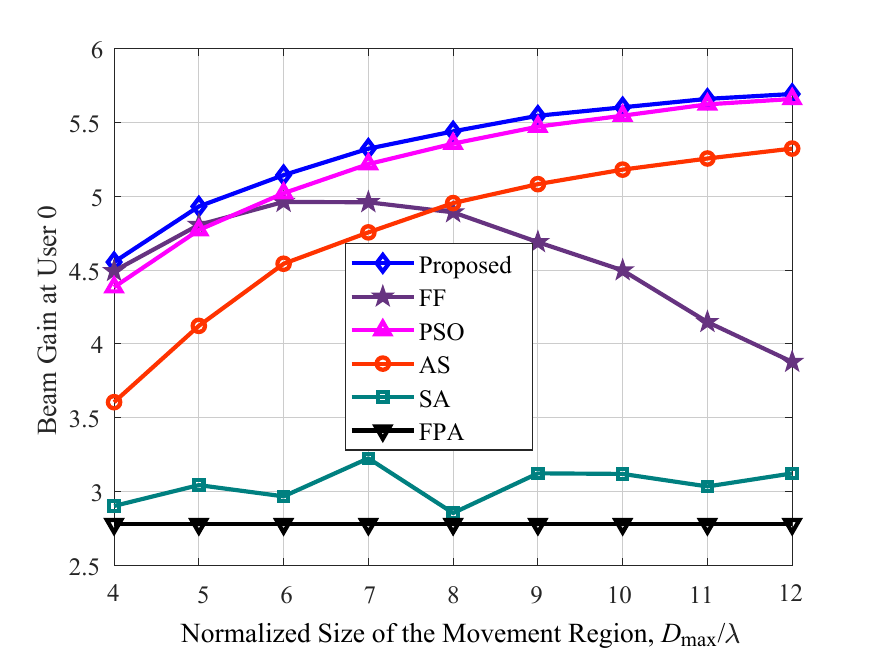}
	\caption{Beam gain at user 0 versus size of the movement region in Scenario 1.}
	\label{Fig_S1_TxRegion}
    \vspace{-9pt}
\end{figure}
In Fig.~\ref{Fig_S1_TxRegion}, we show the beam gain at user 0 versus the normalized size of the movement region $D_{\max}/\lambda$ with $K=3$. It is observed that the proposed scheme yields a higher beam gain at user 0 compared to other benchmark schemes. Moreover, by enlarging the size of the movement region, the beam gain by the proposed scheme increases and gradually approaches the full beam gain (i.e., 6). This improvement arises since a larger movable region enables the MAs to explore more favorable positions, which reduces the spatial correlation of the steering vectors for the target user and the undesired users. In contrast, the beam gains by the SA and FPA benchmarks remain approximately constant as the size of the movement region increases. In particular, the performance of the SA benchmark may slightly degrade as $D_{\max}$ increases due to the uncontrollable grating-lobe effects. {It is also observed that the performance of the far-field design degrades as the normalized movement region size $D_{\max}/\lambda$ increases, indicating its inability to utilize the expanded antenna aperture for enhanced beamforming.}

\begin{figure}[t]
	\centering
	\captionsetup{justification=raggedright,singlelinecheck=false}
	\includegraphics[width=7.6cm]{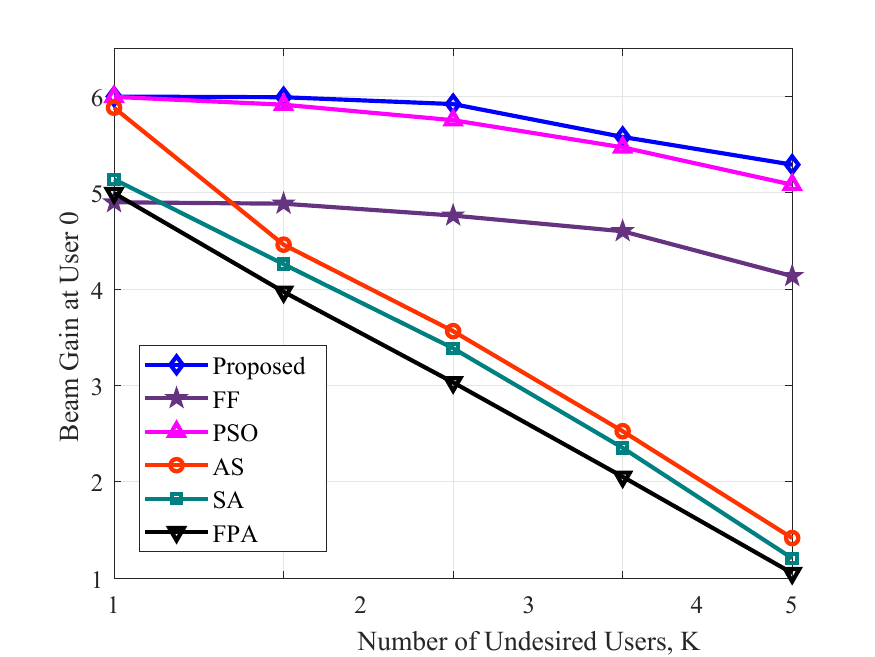}
	\caption{Target user beamforming gain versus different number of nulling users in Scenario 1.}
	\label{Fig_S1_UserNum}
    \vspace{-6pt}
\end{figure}
In Fig.~\ref{Fig_S1_UserNum}, we plot the beam gain at user 0 versus the number of undesired users $K$. It is observed that the beam gains by all schemes decrease as $K$ increases, due to the more stringent constraints for beam nulling. The proposed scheme is observed to outperform all benchmark schemes, and the performance gap increases as $K$ grows. This is because manipulating the spatial correlation of steering vectors becomes increasingly difficult for a larger value of $K$. Conventional AS and PSO algorithms may lack the capability to achieve this purpose. These results underscore the importance of MAs for effective beam nulling in the presence of many undesired directions.\vspace{-6pt}
\begin{figure}[t]
	\centering
	\captionsetup{justification=raggedright,singlelinecheck=false}
	\includegraphics[width=7.6cm]{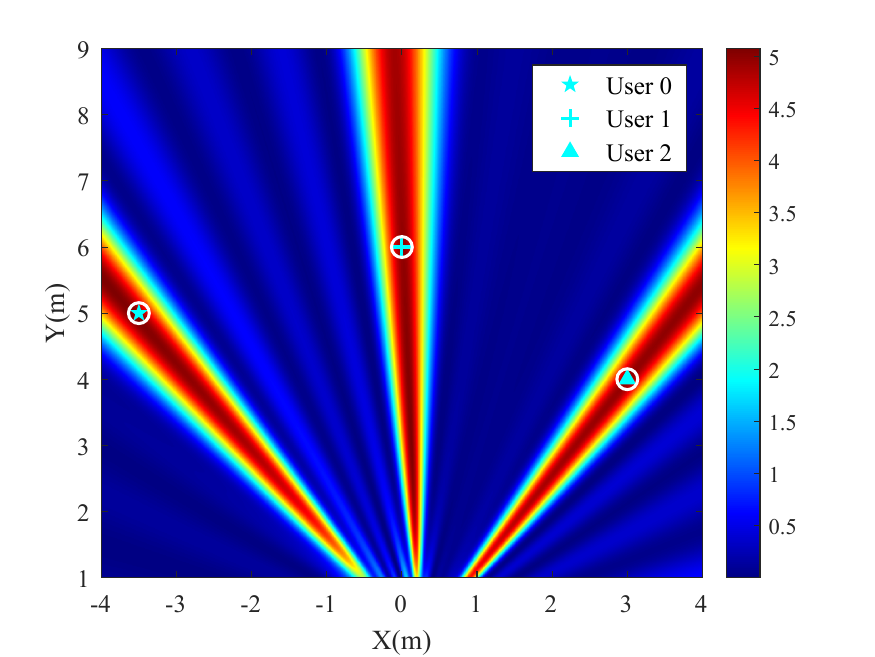}
	\caption{Distribution of beam gain in Scenario 2.}
	\label{Fig_S2_Heatmap}
    \vspace{-6pt}
\end{figure}
\begin{figure}[!t]
\centering
\includegraphics[width=7.6cm]{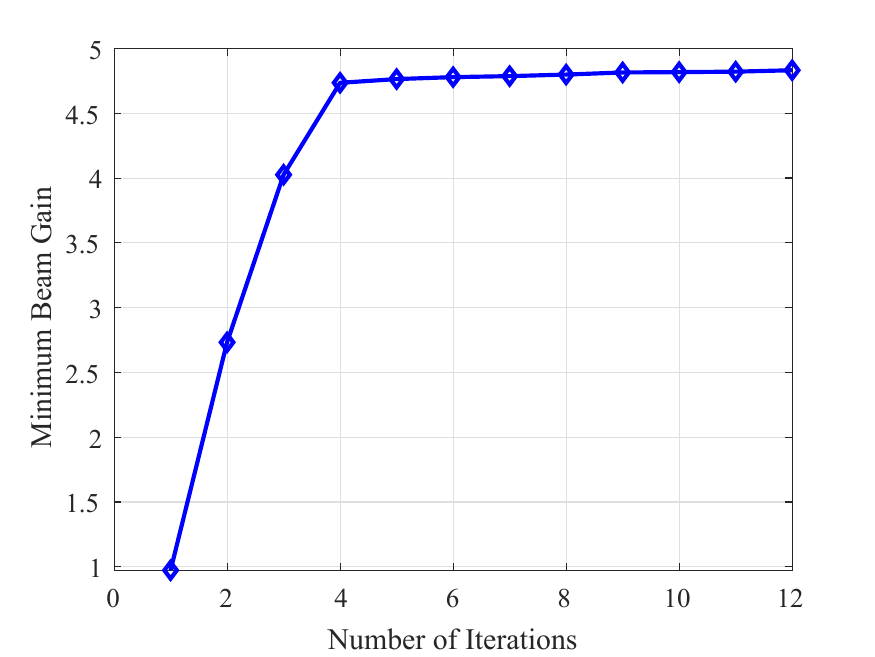}
\caption{Convergence behavior of the proposed AO algorithm in Scenario 2.}
\vspace{-6pt}
\label{Fig_convergence}
\end{figure}

\subsection{Scenario 2: Multi-Beam Forming}
In this subsection, we consider the multi-beam forming scenario. First, we plot in Fig.~\ref{Fig_S2_Heatmap} the distribution of beam gain with $K+1=3$ users and $N=6$ MAs. The three users are located from left to right at $(R_0, \theta_0)=(6.10\;{\text{m}}, 2.18\;{\text{rad}})$, $(R_1, \theta_1)=(6\;{\text{m}}, 1.57\;{\text{rad}})$, and $(R_2, \theta_2)=(5\;{\text{m}}, 0.93\;{\text{rad}})$, respectively. It is observed that the beam gains achieved at these users are $4.96, 4.96$ and $4.95$, respectively. Notably, the majority of the beam energy is uniformly concentrated around the user locations, thereby ensuring reliable communication performance. This demonstrates the effectiveness of the proposed algorithm in generating high-gain beams at specific locations in the near field. 

{In Fig.~\ref{Fig_convergence}, we show the convergence behavior of the proposed AO algorithm for the multi-beam forming scenario with $N=6$ and $K+1=3$ users. The output of the algorithm is observed to stabilize after any 5 iterations. This confirms the convergence of our proposed algorithm, which is consistent with our theoretical analysis in Section \ref{Sec4}.}

\begin{figure}[!t]
	\centering
	\captionsetup{justification=raggedright,singlelinecheck=false}
	\includegraphics[width=7.6cm]{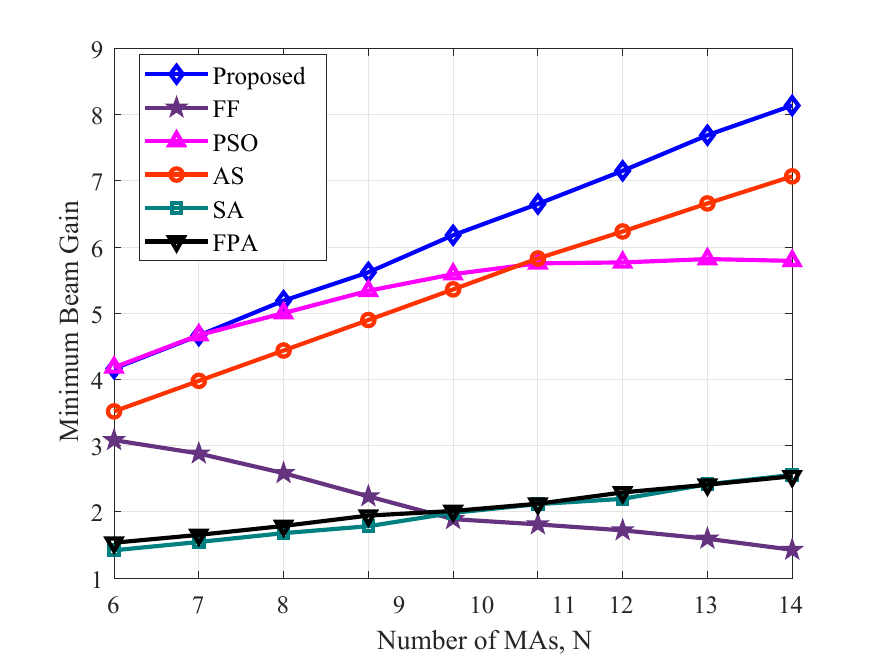}
	\caption{Max-min beamforming gain versus the number of antennas in Scenario 2.}
	\label{Fig_S2_TxNums}
    \vspace{-6pt}
\end{figure}
Next, we plot in Fig.~\ref{Fig_S2_TxNums} the max-min beam gain versus the number of MAs with $K+1=3$ users and $D_{\max} = 1.5N\lambda$. It is observed that the proposed scheme significantly outperforms all the benchmark schemes. For example, when $N=14$, the max-min beam gain achieved by our proposed algorithm is $8.14$, while that achieved by the FPA benchmark is only $2.53$. This improvement is attributed to the more favorable spatial correlations among the steering vectors achieved by the optimized antenna positions. It is also noteworthy that the performance of PSO shows negligible improvement for $N \ge 10$ and even falls behind that of the AS scheme. This highlights the advantage of the proposed sequential update algorithm, particularly in scenarios with a large $N$. {Furthermore, the minimum beam gain of the far-field design is observed to be significantly lower than that of the near-field design, especially as $N$ increases.}

In Fig.~\ref{Fig_S2_TxRegion}, we illustrate the max-min beam gain versus the normalized size of the movement region with $K=2$. It is observed that as $D_{\max}$ increases from $4\lambda$ to $12\lambda$, the max-min beam gain achieved by the proposed scheme increases monotonically and consistently outperforms those of the benchmark schemes, similar to the trend observed in the beam-nulling scenario (see Fig.~\ref{Fig_S1_TxRegion}). In contrast, although the SA benchmark is capable of generating grating lobes, its max-min beam gain is even lower than that of the FPA benchmark. This suggests that SA lacks sufficient degrees of freedom to effectively steer grating lobes toward the user locations.
\begin{figure}[t]
	\centering
	\captionsetup{justification=raggedright,singlelinecheck=false}
	\includegraphics[width=7.6cm]{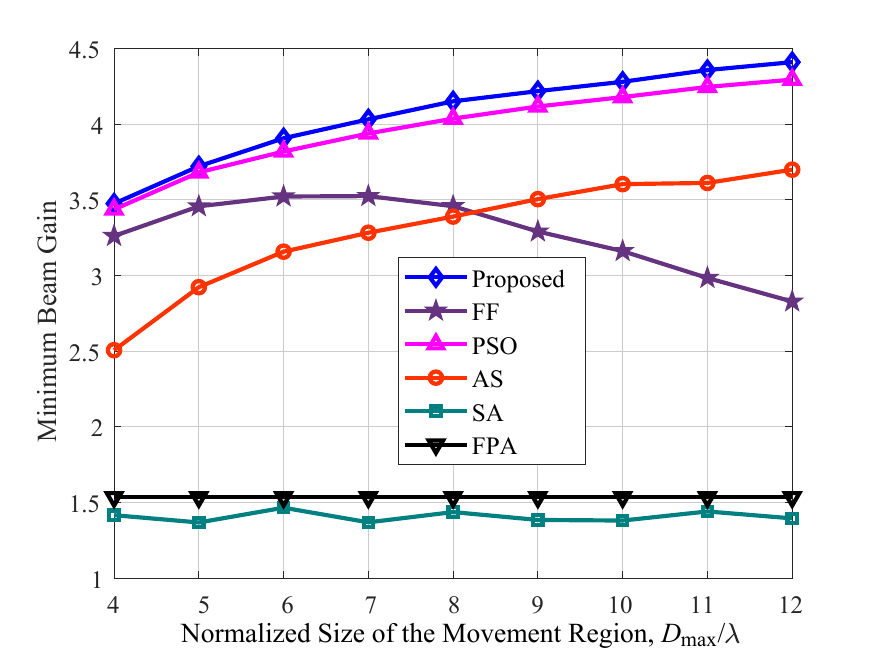}
	\caption{Max-min beamforming gain versus the size of moving range in Scenario 2.}
	\label{Fig_S2_TxRegion}
    \vspace{-6pt}
\end{figure}
\begin{figure}[t]
	\centering
	\captionsetup{justification=raggedright,singlelinecheck=false}
	\includegraphics[width=7.6cm]{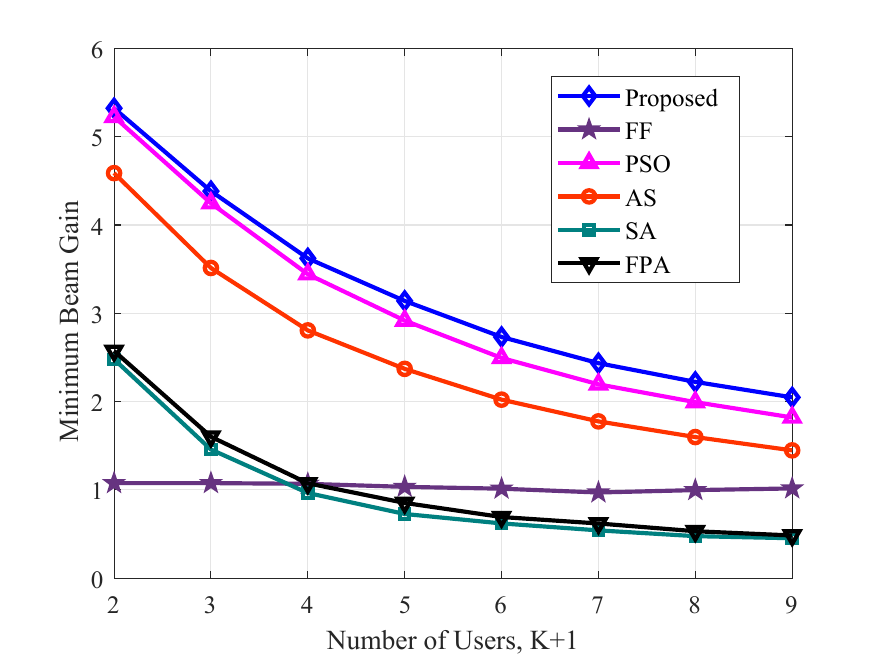}
	\caption{Max-min beamforming gain versus the number of users in Scenario 2.}
	\label{Fig_S2_UserNum}
    \vspace{-6pt}
\end{figure}
\begin{figure}[!t]
	\centering
	\subfigure[Optimized MA positions in scenario 1]{\includegraphics[width=7.6cm]{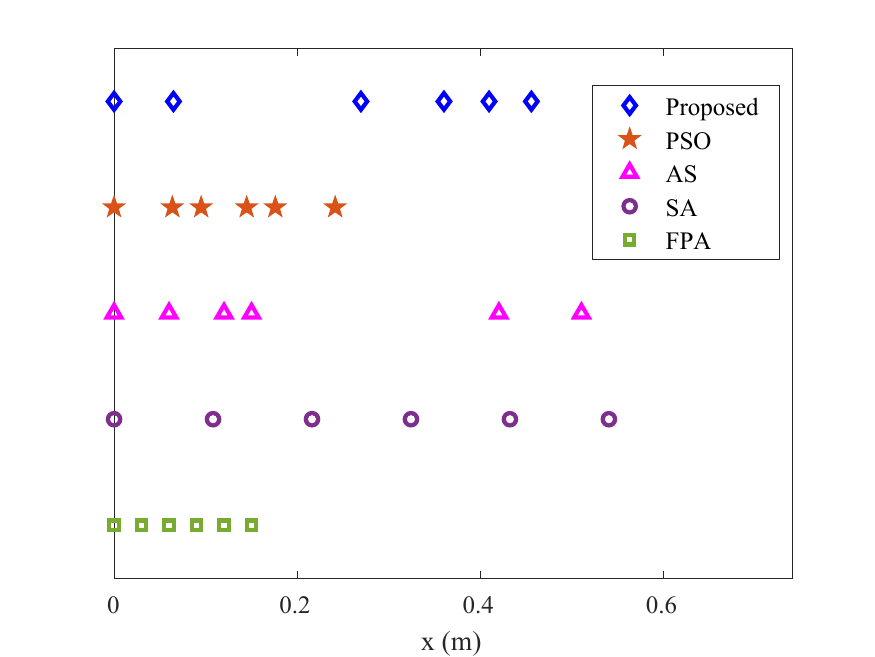}\label{fig_MAlayout_ZF}}
	\subfigure[Optimized MA positions in scenario 2]{\includegraphics[width=7.6cm]{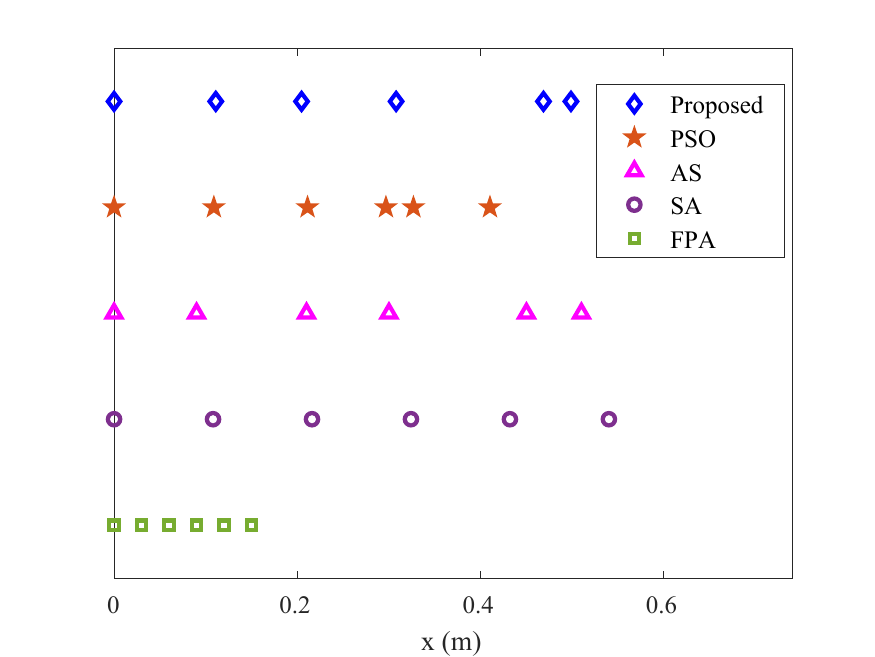}\label{fig_MAlayout_GL}}
	\caption{Optimized 1D MA positions.}
	\label{fig_MApositions}
    \vspace{-6pt}
\end{figure}

In Fig.~\ref{Fig_S2_UserNum}, we plot the max-min beam gain versus the number of users, $K+1$. It is observed that the max-min beam gain decreases with $K$. This is because adding more users imposes additional constraints on the beamforming and APV optimization. Hence, the APV optimization needs to better balance the beam gain across all users rather than concentrating the energy on a single direction. In addition, our proposed scheme consistently outperforms other benchmark schemes. Notably, even with a relatively large number of users (e.g., $K=8$), our proposed scheme can still achieve substantial performance improvement (about fourfold higher beamforming gain) compared to the conventional FPA scheme. This demonstrates the effectiveness of leveraging MAs even in the case of a large number of grating lobes.\\
{\indent  Fig.~\ref{fig_MApositions} shows the optimized MA positions in the two considered scenarios. In Scenario 1 (Fig.~\ref{fig_MAlayout_ZF}), the MAs tend to cluster more compactly within the movement region to achieve precise phase alignment for interference suppression. In contrast, for Scenario 2 (Fig.~\ref{fig_MAlayout_GL}), the MAs are distributed more sparsely to generate simultaneous grating lobes toward multiple user directions. This contrast highlights how different design objectives lead to fundamentally different spatial configurations: compact arrangements for precise nulling versus dispersed layouts for multi-beam coverage. These observations provide intuitive understanding of how antenna repositioning enables flexible beamforming: by adjusting spatial positions, MAs can manipulate the phase relationships among array elements to achieve desired radiation patterns.}

\subsection{Effects of Antenna Position Errors}
In this subsection, we present numerical results to validate our derived analytical results in the case with antenna position errors. The number of MAs is $N=6$. In the beam-nulling scenario, the desired user (user 0) is positioned at $(R_0, \theta_0) = (5\;{\text{m}}, 0.93\;{\text{rad}})$. We consider $K=3$ undesired users located at $(R_1, \theta_1) = (5\;{\text{m}}, 2.21\;{\text{rad}})$, $(R_2, \theta_2) = (6.08\;{\text{m}}, 1.74\;{\text{rad}})$,and $(R_3, \theta_3) = (4.47\;{\text{m}}, 0.46\;{\text{rad}})$. In the multi-beam forming scenario, we consider $K+1=2$ users located at $(R_0, \theta_0) = (8.94\;{\text{m}}, 2.03\;{\text{rad}})$, and $(R_1, \theta_1) = (7.61\;{\text{m}}, 1.16\;{\text{rad}})$. 

\begin{figure}[t]
	\centering
	\subfigure[Without antenna position error]{\includegraphics[width=7.6cm]{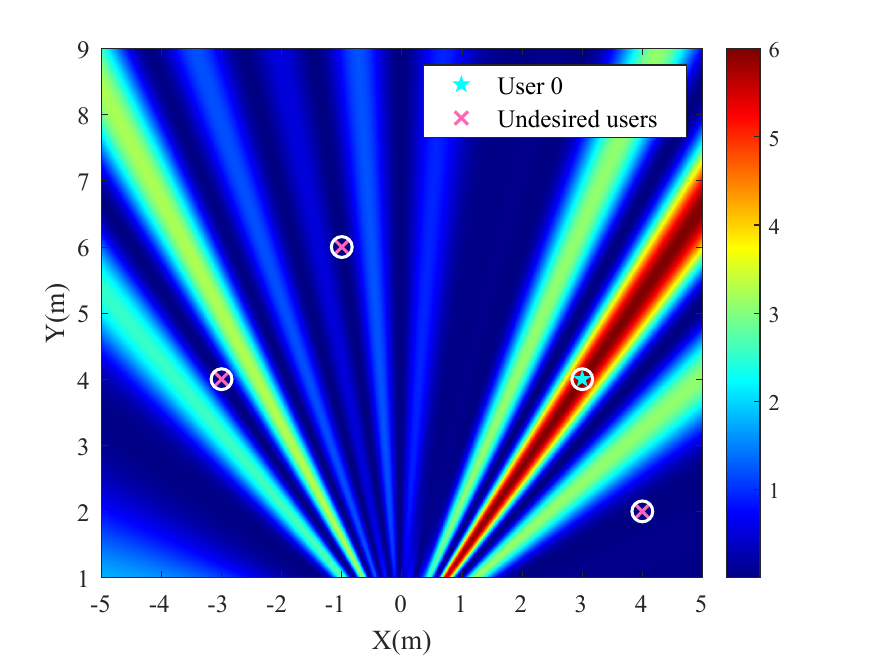}\label{fig:zf_ideal}}
	\subfigure[With antenna position error]{\includegraphics[width=7.6cm]{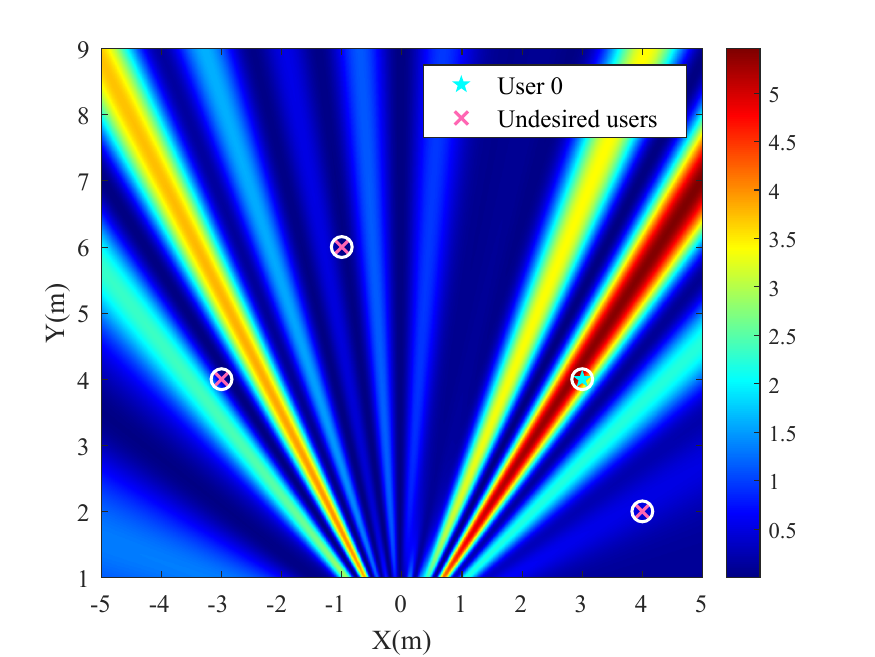}\label{fig:zf_error}}
	\caption{Distribution of the worst-case beam gain in the beam-nulling scenario (Scenario 1).}
	\label{fig:zf_beam_pattern}
\end{figure}
\begin{figure}[t]
	\centering
	\subfigure[Without position error]{\includegraphics[width=7.6cm]{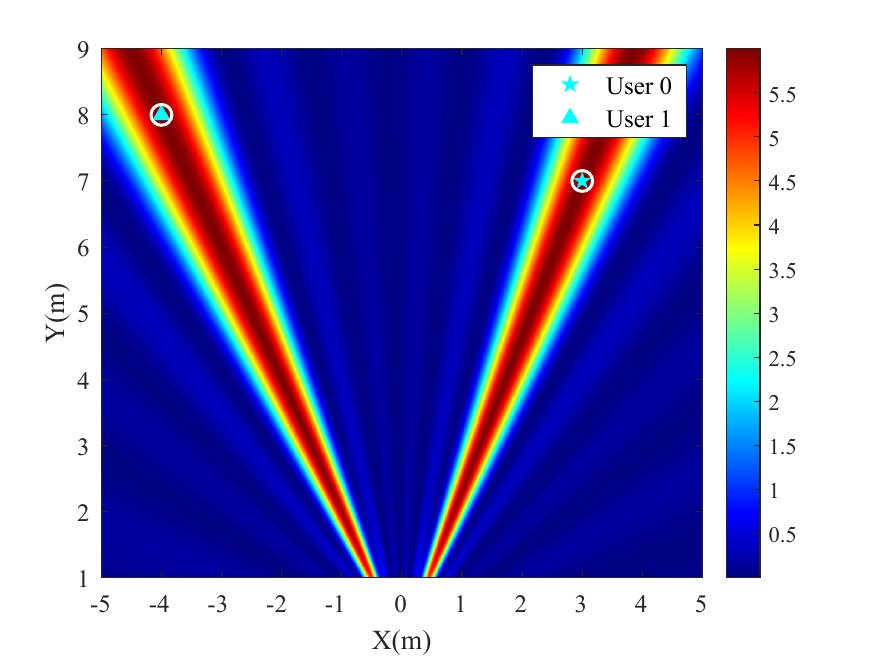}\label{fig:gl_ideal}}
	\subfigure[With position error]{\includegraphics[width=7.6cm]{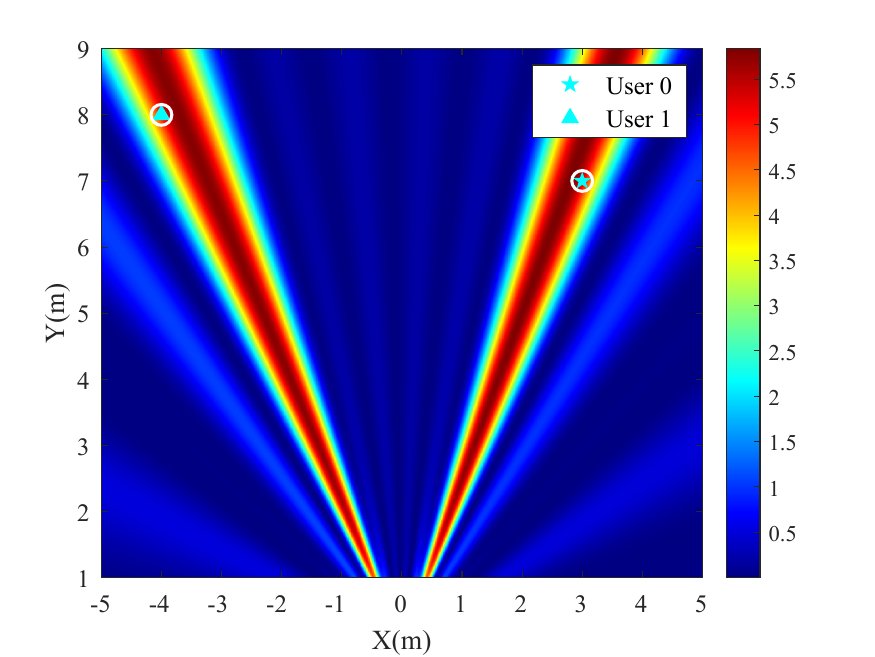}\label{fig:gl_error}}
	\caption{Distribution of the worst-case beam gain in the multi-beam forming scenario (Scenario 2).}
	\label{fig:gl_beam_pattern}
    \vspace{-6pt}
\end{figure}
First, Figs. \ref{fig:zf_beam_pattern} and \ref{fig:gl_beam_pattern} depict the distribution of the worst-case beam gain in Scenarios 1 and 2 (see \eqref{eq_squared_magnitude_expansion} and \eqref{eq_best_case_gain_k_re}), respectively. The maximum position error is set as $\epsilon = 0.009$ m. It is observed from Fig. \ref{fig:zf_ideal} that without antenna position errors, the target user 0 receives a full beam gain, while effective signal nulling is achieved at the three undesired users. However, when antenna position errors are introduced, it is observed from Fig. \ref{fig:zf_error} that the signal nulls are significantly disrupted, leading to increased beam gain toward the undesired users. Similar observations can also be made from Fig.\,\ref{fig:gl_beam_pattern} in Scenario 2, where the presence of antenna position errors leads to reduced beam gains at the two users. Interestingly, user 0 is observed to experience less beam-gain degradation in both scenarios. These observations demonstrate the detrimental impact of antenna position errors on beam nulling and multi-beam forming performance, especially for users not aligned with the main beam. 

\begin{figure}[t]
	\centering
	\includegraphics[width=7.4cm]{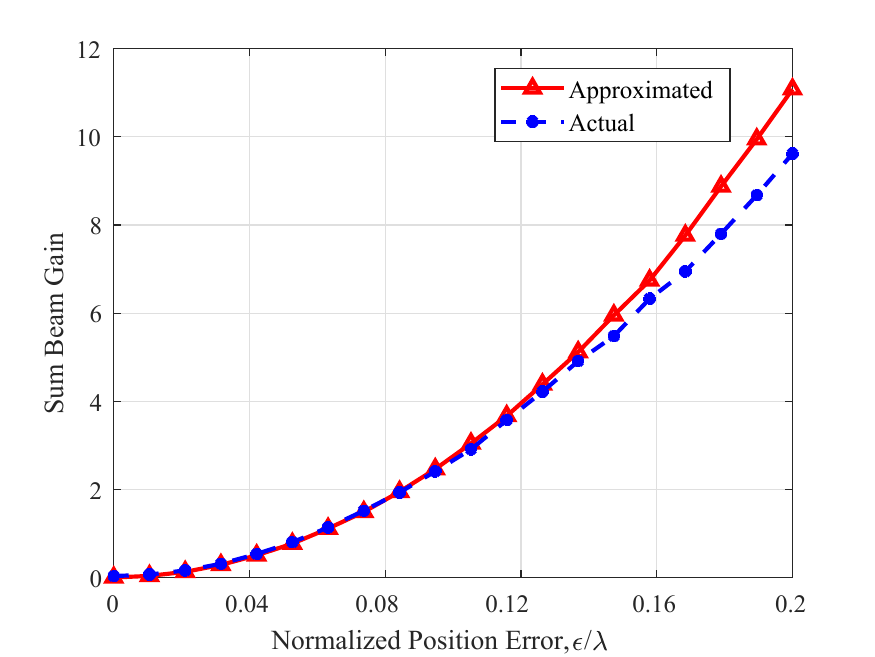} 
	\caption{Actual and approximated sum beam gain versus position error in Scenario 1}
	\label{fig:zf_metrics_comparison}
    \vspace{-6pt}
\end{figure}
\begin{figure}[t]
	\centering
	\includegraphics[width=7.4cm]{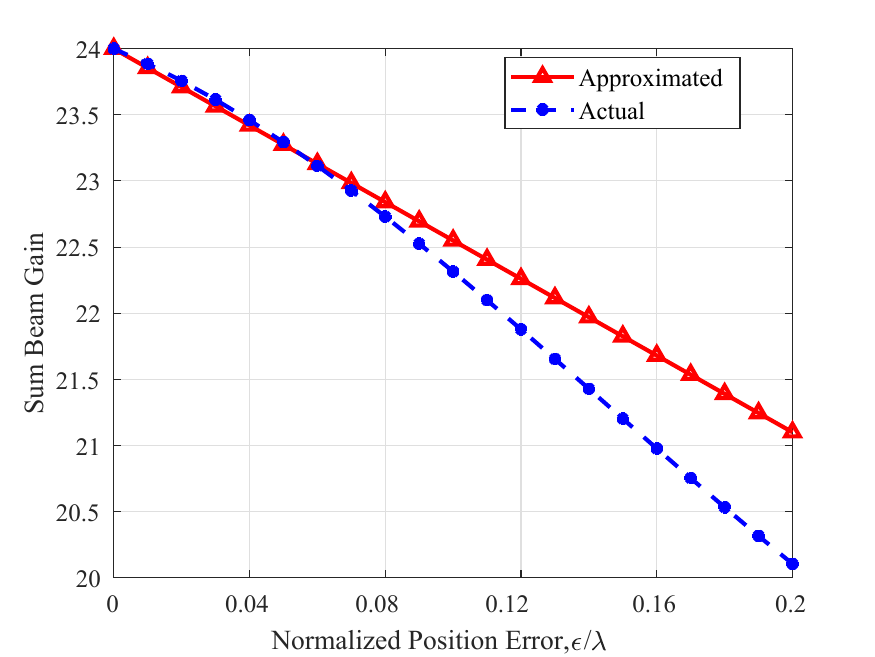} 
	\caption{Actual and approximated sum beam gain versus position error in Scenario 2}
	\label{fig:gl_sumgain_comparison}
    \vspace{-6pt}
\end{figure}
To further quantify the impact of antenna position errors and validate our analytical results, Figs. \ref{fig:zf_metrics_comparison} and Fig. \ref{fig:gl_sumgain_comparison} plot the worst-case sum beam gain at the $K$ users versus the maximum position error (normalized by the wavelength), i.e., $\epsilon/\lambda$, in Scenarios 1 and 2, respectively. To evaluate the accuracy of the proposed approximation method, we also show the actual sum beam gain achieved by the optimized $\bs{\Delta d}$ for both scenarios in Figs. \ref{fig:zf_metrics_comparison} and \ref{fig:gl_sumgain_comparison}, by substituting it into the objective functions of problems \eqref{eq_problem_zf} and \eqref{eq_problem_gl}, respectively. It is observed from Fig. \ref{fig:zf_metrics_comparison} that the worst-case sum beam gain increases with the normalized position error, implying that larger position errors lead to more significant interference leakage to the $K$ undesired users. It is also observed that the proposed method achieves high approximation accuracy when the maximum position error is small, e.g., $\epsilon/\lambda \leq 0.15$. This is expected, as the Taylor expansion provides accurate approximations under small perturbations. Furthermore, it is observed from Fig.\,\ref{fig:gl_sumgain_comparison} that increasing the position error degrades the worst-case sum beam gain in Scenario 2 for multi-beam forming, as expected. The proposed approximation method maintains high accuracy in this scenario as well (with the maximum difference less than 1 dB), consistent with the observations made from Fig. \ref{fig:zf_metrics_comparison}.

\begin{figure}[t]
	\centering
	\includegraphics[width=7.4cm]{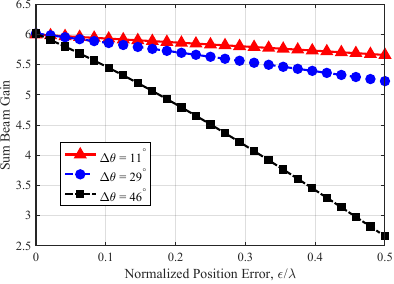} 
	\caption{Worst-case sum beam gain versus normalized position error under different angular separations ($\Delta\theta$) in Scenario 1.}
	\label{fig:gl_sumgain_diffangluarsep}
    \vspace{-6pt}
\end{figure}

{Fig. \ref{fig:gl_sumgain_diffangluarsep} shows the worst-case sum beam gain versus normalized position error under different angular separations ($\Delta\theta = 11^\circ, 29^\circ, 46^\circ$) with $K+1=2$ users and $N=6$ MAs, where $\Delta\theta$ represents the angular difference between the target user and the undesired user. It is observed that at $\epsilon/\lambda = 0.3$, the wide angular separation  ($\Delta\theta = 46^\circ$) leads to a gain reduction of approximately 1.8 compared to the ideal case with $\epsilon/\lambda=0$. In contrast, the narrow angular separation ($\Delta\theta = 11^\circ$) suffers only about 0.2 beam-gain loss. These results demonstrate that larger angular spreads significantly amplify the system's sensitivity to antenna position errors, validating our analysis presented at the end of Section V-B.}

\vspace{-6pt}

\section{Conclusion}
In this paper, we investigated beam nulling and multi-beam forming in the near-field by employing a linear MA array. Specifically, we aimed to jointly optimize the transmit beamforming and APV to maximize the beam gain toward a target direction under the ZF beamforming in the scenario of beam nulling. Whereas in the scenario of multi-beam forming, we jointly optimized the transmit beamforming and APV to maximize the minimum beam gain among all directions. Under the ALMR assumption, we showed that using MRT toward a target user is able to achieve nulls and full beam gains at other users in the near field. In the general case, efficient algorithms were proposed to obtain high-quality APV and beamforming solutions. Next, the effects of antenna position errors were also analyzed by deriving the worst-case sum beam gain in the above two scenarios. Our analytical results demonstrated that the carrier frequency and user distribution can both influence the extent of performance degradation caused by antenna position errors. Finally, simulation results verified the effectiveness of our proposed optimization algorithms and approximation method, thus offering a new array signal processing approach for steering spatial correlation in the near field. 

{While this work focuses on 1D linear arrays for analytical tractability, extending the results to other MA architectures (e.g., 2D planar arrays, circular arrays, or ultra-massive MAs) represents promising future directions. Furthermore, it is also interesting to consider antenna position optimization and performance analysis incorporating both beam nulling and multi-beam forming. Our current results are expected to provide foundational insights and optimization frameworks for these future directions.}
\vspace{-6pt}

\bibliographystyle{IEEEtran}
\bibliography{ref_shun}

\end{document}